\documentclass[10pt,a4paper]{article}
\usepackage{amsmath}
\usepackage{amsfonts}
\usepackage{amssymb}
\usepackage[dvips]{graphicx}
\usepackage{bm}

\begin{document}

\title{\textbf{Dynamical formation and evolution of\\(2+1)-dimensional charged black holes}}
\author{\textsc{Dong-il Hwang}$^{a}$\footnote{dongil.j.hwang@gmail.com},\;\; \textsc{Hongbin Kim}$^{b}$\footnote{hongbin@yonsei.ac.kr}\;\;
and \textsc{Dong-han Yeom}$^{a,c,d}$\footnote{innocent.yeom@gmail.com}\\
\textit{$^{a}$\small{Department of Physics, KAIST, Daejeon 305-701, Republic of Korea}}\\
\textit{$^{b}$\small{Department of Physics, College of Science, Yonsei University, Seoul 120-749, Republic of Korea}}\\
\textit{$^{c}$\small{Center for Quantum Spacetime, Sogang University, Seoul 121-742, Republic of Korea}}\\
\textit{$^{d}$\small{Research Institute for Basic Science, Sogang University, Seoul 121-742, Republic of Korea}}}
\maketitle

\begin{abstract}
In this paper, we investigate the dynamical formation and evolution of $2+1$-dimensional charged black holes.
We numerically study dynamical collapses of charged matter fields in an anti de Sitter background and note the formation of black holes using the double-null formalism. Moreover, we include re-normalized energy-momentum tensors assuming the $S$-wave approximation to determine thermodynamical back-reactions to the internal structures.
If there is no semi-classical effects, the amount of charge determines the causal structures. If the charge is sufficiently small, the causal structure has a space-like singularity. However, as the charge increases, an inner Cauchy horizon appears. If we have sufficient charge, we see a space-like outer horizon and a time-like inner horizon, and if we give excessive charge, black hole horizons disappear. We have some circumstantial evidences that weak cosmic censorship is still satisfied, even for such excessive charge cases. Also, we confirm that there is mass inflation along the inner horizon, although the properties are quite different from those of four-dimensional cases.
Semi-classical back-reactions will not affect the outer horizon, but they will affect the inner horizon. Near the center, there is a place where negative energy is concentrated. Thus, charged black holes in three dimensions have two types of curvature singularities in general: via mass inflation and via a concentration of negative energy. Finally, we classify possible causal structures.
\end{abstract}

\newpage

\tableofcontents

\newpage

\section{\label{sec:int}Introduction}

Traditionally, people regarded that three-dimensional models are good toy models with which to study the basic nature of pure gravity and quantum gravity \cite{Carlip:1998uc}. First, there may be a qualitative relation between a cosmic string in four dimensions and a point source in three dimensions \cite{Deser:1983tn}. Second, three-dimensional gravity is simpler than the four-dimensional version, making it useful for obtaining some intuitions about quantum gravity using gauge theoretical correspondences \cite{Witten:1988hc}. However, given that this is too simple, it was doubtful that a black hole solution would be found in a three-dimensional background.

The first analytic solution in three-dimensional gravity was found by Banados, Teitelboim, and Zaneli \cite{Banados:1992wn}\cite{Banados:1992gq}. They found the black hole solution in an anti de Sitter background. In their solutions, the negative cosmological constant played a crucial role. These solutions can have mass, charge, and angular momentum. Moreover, the solutions can have the Hawking temperature \cite{Carlip:1994gc}. However, they are thermodynamically stable because the specific heat is positive \cite{Reznik:1991qj}.

The so-called BTZ solutions are useful in the context of holography. According to the AdS/CFT correspondence \cite{Maldacena:1997re}, there is a relationship between the bulk three-dimensional gravity and the conformal field theory at the asymptotic two-dimensional boundary of such a three-dimensional bulk. That is, we may relate the gravitational perturbation on the bulk to the thermal excitation on the boundary field theory. Moreover, it is known that the U-duality transformation maps four or five-dimensional black holes to three-dimensional BTZ black holes \cite{Hyun:1997jv}; here, these black holes, which are connected by the U-duality, have the same number of microstates. Hence, the microscopic derivation of the entropy can be replaced by the study on the corresponding derivation for the BTZ black hole \cite{Strominger:1997eq}. In these senses, three-dimensional black holes are more important than just an artificial toy model.

In this paper, we will begin to study the dynamics of $2+1$-dimensional charged black holes, including gravitational collapses and back-reactions, using the semi-classical effects of Hawking radiation. We solved time-dependent Einstein and field equations using the double-null formalism \cite{doublenull}\cite{Hong:2008mw}\cite{Hong:2008mw_2}. Of course, it is too ambitious to extend this method to holography or string theory in this paper. However, if we establish a method with which to study dynamical black holes in an anti de Sitter background, it will eventually become possible to study the time-dependent correspondences between the bulk and boundary someday.

Thus far, there is less reference to compare our calculations on the dynamic formation of three-dimensional charged black holes (e.g., \cite{Crisostomo:2003xz}), whereas there is some literature on the gravitational collapses of neutral matter \cite{Gutti:2005pn}, especially in the context of critical collapses \cite{Alcubierre:1999ex}. On the other hand, in four dimensions, there are numerous studies of gravitational collapses and evaporations \cite{doublenull}\cite{Hong:2008mw}. One important feature in four-dimensional dynamical charged black holes is the existence of mass inflation \cite{Poisson:1989zz}. Mass inflation is a phenomenon in which the local mass function of an observer near the Cauchy horizon increases exponentially. If it is true, then the internal structure of charged black holes cannot be described by the Reissner-Nordstr\"{o}m solution \cite{RN} and the internal structure should be significantly modified \cite{Bonanno:1994ma}. This was directly confirmed by numerical simulations, and there are more related works \cite{doublenull}\cite{Hong:2008mw}. Three-dimensional gravity will be similar as regards this point, as it is expected that there is mass inflation in general in charged black holes \cite{Poisson:1997my}. However, there are some expectations that three-dimensional mass inflation is different from four-dimensional cases \cite{Cai:1995ib}. Moreover, because black holes are thermodynamically stable in three-dimensions, it will be interesting to determine what the semi-classical back-reactions are in the internal structures of black holes (there are discussions on the semi-classical effects in the static limit \cite{Steif:1993zv}).

This paper is organized as follows:
in Section~\ref{sec:pre}, we summarize analytic solutions for three-dimensional charged black holes and discuss the mass inflation issue. The paper will then provide the theoretical motivations to study three dimensions.
In Section~\ref{sec:mod}, we describe our model of the numerical simulations, the implementation to the double-null formalism, and comment on the initial conditions and dimensional analysis.
In Section~\ref{sec:dyn}, we report the results. First, we report results for cases without semi-classical effects. Second, we include semi-classical effects and observe how they modify causal structures.
Finally, in Section~\ref{sec:con}, we conclude and comment on causal structures and cosmic censorship issues.

\section{\label{sec:pre}Preliminaries}

In this section, first we briefly review the analytic and static charged black hole solutions of three dimensions. Second, we summarize thermodynamic properties and internal stability issues via mass inflation regarding their solutions.

\subsection{Static solutions: causal structures and thermodynamics}

\subsubsection{Three dimensions: BTZ solutions}

For three dimensions, the first simple black hole solution in the anti de Sitter background is known to be the BTZ solution (for non-rotating cases) \cite{Banados:1992wn}. For the Einstein-Maxwell action
\begin{eqnarray}\label{eqn:3d}
S = \frac{1}{16 \pi} \int dx^{3} \sqrt{-g} \left(R - 2 \Lambda - F_{\mu\nu}F^{\mu\nu}\right),
\end{eqnarray}
the solution becomes
\begin{eqnarray}
ds_{\mathrm{BTZ}}^{2} = - N^{2} dt^{2} + N^{-2} dr^{2} + r^{2} d \varphi^{2},
\end{eqnarray}
where
\begin{eqnarray}
N^{2}(r) = -M + \frac{r^{2}}{l^{2}} - Q^{2} \ln \frac{r^{2}}{l^{2}},
\end{eqnarray}
$\Lambda = - 1/l^{2}$, $Q$ is electric charge, and $M$ is interpreted by the mass.

\begin{figure}
\begin{center}
\includegraphics[scale=0.5]{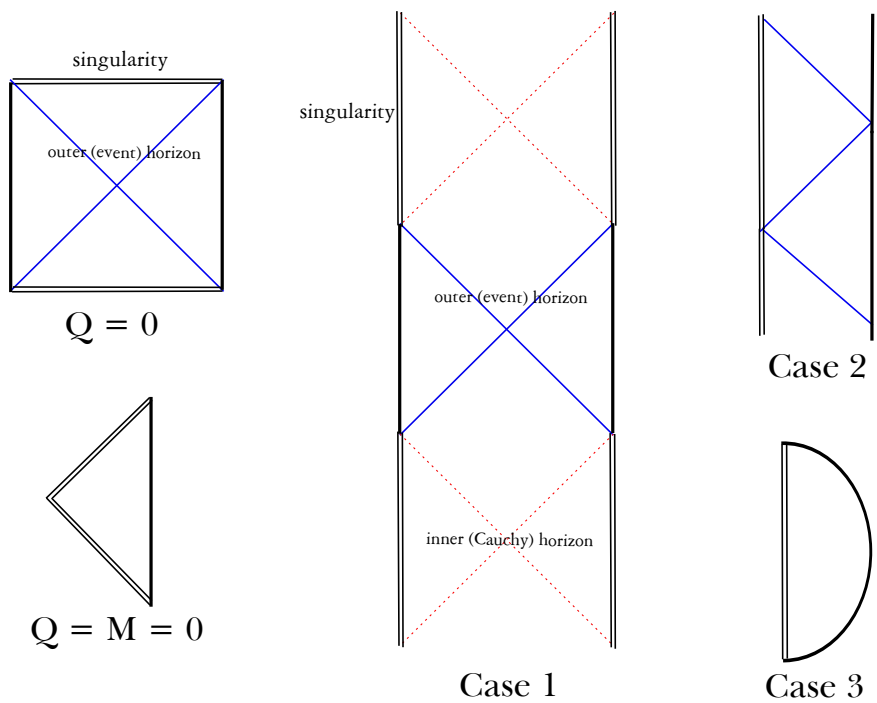}
\caption{\label{fig:3D}Penrose diagrams of BTZ solutions for $Q=0$ and $M>0$, $M=Q=0$, Case~$1$ (there are two horizons), Case~$2$ (there is one horizon), and Case~$3$ (there is no horizon but a time-like singularity).}
\end{center}
\end{figure}

If $Q=0$, then we obtain a neutral black hole solution with $r_{+} = l\sqrt{M}$. Thus although this is in the $M=0$ limit, the center still contains a singularity ($Q=M=0$ in Figure~\ref{fig:3D}). The proper anti de Sitter limit is obtained by the $M=-1$ limit:
\begin{eqnarray}
ds^{2} = - \left( 1 + \frac{r^{2}}{l^{2}} \right) dt^{2} + \left( 1 + \frac{r^{2}}{l^{2}} \right)^{-1} dr^{2} + r^{2} d \varphi^{2}.
\end{eqnarray}

If $Q\neq0$, then there can be two horizons: the outer horizon $r_{+}$ and the inner horizon $r_{-}$. If $M>1$, then there are always two horizons with $r_{+}>l\sqrt{M}$ and $r_{-}<l\sqrt{M}$ (Case~$1$ in Figure~\ref{fig:3D}). However, if $M<1$, then the number of horizons depends on $Q$: for a small $Q$ limit, there can be two horizons $l\sqrt{M} > r_{+} > r_{-}$ (Case~$1$ in Figure~\ref{fig:3D}); for a large $Q$ limit, there is no horizon (Case~$3$ in Figure~\ref{fig:3D}); and the extreme limit, $r_{\pm} = Q l$ and $M = Q^{2} (1 - \ln Q^{2})$ (Case~$2$ in Figure~\ref{fig:3D}). In the neutral limit, the singularity $r=0$ is space-like ($Q=0$ in Figure~\ref{fig:3D}); in the charged limit, the singularity is time-like (except $Q=M=0$). Therefore, it is not difficult to draw Penrose diagrams in the static limit \cite{Banados:1992gq}.

The Hawking temperature $T_{\mathrm{BTZ}}$ becomes
\begin{eqnarray}
T_{\mathrm{BTZ}} = \frac{1}{4\pi} \left. \frac{dN^{2}}{dr}\right|_{r_{+}} = \frac{r_{+}}{2 \pi l^{2}} \left( 1-\frac{Q^{2}l^{2}}{r_{+}^{2}} \right),
\end{eqnarray}
and $T_{\mathrm{BTZ}}$ is positive or zero because $r_{+}$ is greater than the extreme limit size $Ql$ \cite{Carlip:1998uc}\cite{Carlip:1994gc}. The entropy $S_{\mathrm{BTZ}}$ becomes
\begin{eqnarray}
S_{\mathrm{BTZ}} = \frac{A_{\mathrm{BTZ}}}{4} = \frac{\pi r_{+}}{2}.
\end{eqnarray}
However, the specific heat $C_{\mathrm{BTZ}}$ becomes
\begin{eqnarray}
C_{\mathrm{BTZ}} = \frac{\partial S_{\mathrm{BTZ}}}{\partial T_{\mathrm{BTZ}}} \propto \left( 1 + \frac{Q^{2}l^{2}}{r_{+}^{2}} \right)^{-1},
\end{eqnarray}
and hence the black hole is in thermal equilibrium and the outer horizon does not shrink \cite{Reznik:1991qj}. Therefore, when we consider the dynamical formation of a black hole, the outer horizon should be always space-like or null.

Note that in the double-null coordinates,
\begin{eqnarray}
ds_{\mathrm{BTZ}}^{2} = - \alpha^{2} du dv + r^{2} d \varphi^{2},
\end{eqnarray}
the Einstein equation for the $vv$ component around the outer apparent horizon $r_{,v}=0$ is
\begin{eqnarray}
\left.G_{vv}\right|_{r_{+}} = -\frac{1}{r} r_{,vv} = 8 \pi T_{vv}.
\end{eqnarray}
Therefore, the outer apparent horizon is space-like or null if and only if $T_{vv} \geq 0$. In other words, for BTZ solutions, even when they are dynamically generated with a Hawking temperature, $T_{vv}$ should be positive or $0$ around the outer apparent horizon.

\subsubsection{Instability of Cauchy horizons}

There are a number of discussions on the instability of Cauchy horizons $r_{-}$ \cite{Poisson:1997my}\cite{Cai:1995ib}\cite{cc}. We can re-write the metric in the following form:
\begin{eqnarray}
ds^{2} = - N^{2} dv^{2} + 2 dv dr + r^{2} d \Omega_{D-1}^{2}.
\end{eqnarray}
Here, $D$ is the space dimensions ($2$ or $3$). Now, without loss of generality, we can ignore the angular part to discuss in-going or out-going null geodesics.

Thus, for an in-falling matter along the in-going null direction (for coordinates $[v, r]$), the energy-momentum tensor components $T_{\alpha\beta}$ can be represented by
\begin{eqnarray}
T_{\alpha\beta} \propto \frac{F(v)}{r^{D-1}}\partial_{\alpha}v \partial_{\beta}v,
\end{eqnarray}
where $F(v)$ is an arbitrary function that refers to the luminosity function \cite{Poisson:1997my}\cite{Cai:1995ib}. At this point, if there is an observer who moves along the out-going null direction and approaches the Cauchy horizon $r_{-}$, then for a null geodesic of the observer $l^{\alpha}$, $(\partial_{\alpha}v) l^{\alpha} = l^{v} = dv/d\eta$ and $dv/d\eta \propto \exp{\kappa_{-}v}$ around the $v\rightarrow \infty$ limit, where $\eta$ is an affine parameter of the observer and $\kappa_{-}$ is the surface gravity of the inner horizon. In conclusion, the out-going observer who approaches the Cauchy horizon measures the energy density for the energy flow that flows along the Cauchy horizon by \cite{Poisson:1997my}\cite{Cai:1995ib}:
\begin{eqnarray}
\rho = T_{\alpha\beta}l^{\alpha}l^{\beta} \propto \frac{F(v)}{r^{D-1}} e^{2 \kappa_{-} v}.
\end{eqnarray}

Note that for four dimensions, for a realistic black hole the behavior of the luminosity function should be polynomial to $F(v) \sim v^{-p}$ \cite{Price}. Although there is no such guidance regarding the luminosity function for three dimensions, if we believe that such a polynomial fall-off is reasonable for three-dimensional cases, then we can conclude that for both three- and four-dimensional cases, any small energy flow along the Cauchy horizon is amplified for an observer who approaches to the Cauchy horizon. This effect is known as \textit{mass inflation}.

Therefore, the inner horizon cannot be stable and there should be strong back-reactions due to high energy density \cite{Bonanno:1994ma}. In many cases, it is believed that the Cauchy horizon should be a curvature singularity \cite{Ori}. Hence, we cannot trust the static solutions to study the inside structures of charged black holes. Therefore, it is necessary to introduce numerical tools to study these structures.

\subsection{Motivation of this paper}

We have two main motivations for this paper.

\begin{enumerate}
\item \textbf{Theoretical interests:} Three-dimension is important. First, it is a toy model of a black string that may be possible in the early universe. Second, as we commented in the introduction, in string theory, there is a holographic correspondence between bulk black holes in an anti de Sitter space and its boundary conformal field theory. One can relate the gravitational perturbation on the bulk to the thermal excitation on the boundary field theory and one may restrict the perturbation by the spherically symmetric scalar sector. Although the effect of the scalar perturbation on the boundary field theory is not yet revealed fully, our study and methodology on dynamical black holes may shed light on the study of the dynamical correspondence between gravity and field theory.

\item \textbf{Numerical interests:} According to previous discussions, charged black holes in three dimensions will have mass inflation around the inner horizon. This means that the inner horizon and internal structures cannot be described by analytical methods. To induce this mass inflation, the black hole should be fully dynamical. Then it is inevitable to use numerical methods.
\end{enumerate}

Therefore, there are various interests to study three-dimensional black holes. To study internal structures, it is inevitable to use the numerical methods. In this paper, we study the dynamical formation of spherical symmetric charged black holes and its back-reactions for three dimensions. We prepare a complex scalar field with the Maxwell field; it is used to form a charged black hole. Then, it will make it possible to understand mass inflation and the possible curvature singularities inside of three-dimensional black holes.

Moreover, we can include Hawking radiation by introducing re-normalized energy-momentum tensors with the $S$-wave approximation. We can check whether the horizon behavior and the energy condition are consistent with the thermodynamics results, even when we include the re-normalized energy-momentum tensors. We can also determine the back-reaction of the re-normalized energy-momentum tensor to the inside structure of the black hole.

\section{\label{sec:mod}Model for three-dimensional gravity}

In this section, we introduce a model for three-dimensional charged black holes. To use numerical simulations, we introduce the double-null formalism and discuss the details of the initial conditions and free parameters for the simulations. Also, we briefly comment on the dimensional analysis of each functions and parameters.

\subsection{\label{sec:The}Theory}

Let us introduce the action in three dimensions by
\begin{eqnarray}
S = \int dx^{3} \sqrt{-g} \left[ \frac{1}{16 \pi} \left( R - 2\Lambda \right) - \frac{1}{2}\left(\phi_{;\mu}+ieA_{\mu}\phi \right)g^{\mu\nu}\left(\overline{\phi}_{;\nu}-ieA_{\nu}\overline{\phi}\right) -\frac{1}{16\pi}F_{\mu\nu}F^{\mu\nu} \right]
\end{eqnarray}
where $R$ is the Ricci scalar, $\phi$ is a complex scalar field with a gauge coupling $e$ and a gauge field $A_{\mu}$, where $F_{\mu\nu}=A_{\nu;\mu}-A_{\mu;\nu}$.

The Einstein equation becomes as follows:
\begin{eqnarray}\label{eq:Einstein}
G_{\mu\nu} + \Lambda g_{\mu\nu} = 8 \pi T_{\mu\nu},
\end{eqnarray}
where the total energy-momentum tensors are
\begin{eqnarray}\label{eq:matter}
T_{\mu\nu} = T^{\mathrm{C}}_{\mu\nu} + \langle \hat{T}^{\mathrm{H}}_{\mu\nu} \rangle.
\end{eqnarray}
The classical part $T^{C}_{\mu\nu}$ is
\begin{eqnarray}\label{eq:T_C}
T^{\mathrm{C}}_{\mu\nu} &=& \frac{1}{2}\left(\phi_{;\mu}\overline{\phi}_{;\nu}+\overline{\phi}_{;\mu}\phi_{;\nu}\right)
\nonumber \\
&& {}+\frac{1}{2}\left(-\phi_{;\mu}ieA_{\nu}\overline{\phi}+\overline{\phi}_{;\nu}ieA_{\mu}\phi+\overline{\phi}_{;\mu}ieA_{\nu}\phi-\phi_{;\nu}ieA_{\mu}\overline{\phi}\right)
\nonumber \\
&& {}+\frac{1}{4\pi}F_{\mu \rho}{F_{\nu}}^{\rho}+e^{2}A_{\mu}A_{\nu}\phi\overline{\phi}+\mathcal{L}g_{\mu \nu}
\end{eqnarray}
and $\langle \hat{T}^{\mathrm{H}}_{\mu\nu} \rangle$ is the re-normalized energy-momentum tensor part to include semi-classical effects, where $\mathcal{L} = - (1/2)\left(\phi_{;\mu}+ieA_{\mu}\phi \right)g^{\mu\nu}\left(\overline{\phi}_{;\nu}-ieA_{\nu}\overline{\phi}\right) -(1/16\pi)F_{\mu\nu}F^{\mu\nu}$.

The field equations for the complex scalar field $\phi$ and the Maxwell field $A_{\mu}$ are as follows:
\begin{eqnarray}
\label{eq:phi}\phi_{;\mu\nu}g^{\mu\nu}+ieA^{\mu}\left(2\phi_{;\mu}+ieA_{\mu}\phi\right)+ieA_{\mu;\nu}g^{\mu\nu}\phi &=& 0,
\\
\label{eq:A}\frac{1}{2\pi}{F^{\nu}}_{\mu;\nu}-ie\phi\left(\overline{\phi}_{;\mu}-ieA_{\mu}\overline{\phi}\right)+ie\overline{\phi}\left(\phi_{;\mu}+ieA_{\mu}\phi\right) &=& 0.
\end{eqnarray}

\subsection{\label{sec:imp}Implementation via the double-null formalism}

We use the double-null coordinates
\begin{eqnarray}\label{eq:doublenull}
ds^{2} = -\alpha^{2}(u,v) du dv + r^{2}(u,v) d\varphi^{2},
\end{eqnarray}
assuming spherical symmetry, where $u$ is the retarded time, $v$ is the advanced time, and $\varphi$ is the angular coordinate.

We follow the notation of \cite{Hamade:1995ce}\cite{doublenull}\cite{Hong:2008mw}\cite{Hong:2008mw_2}: the metric function $\alpha$, the radial function $r$, and a scalar field $s \equiv \sqrt{4\pi} \phi$, and define
\begin{eqnarray}\label{eq:conventions}
h \equiv \frac{\alpha_{,u}}{\alpha},\quad d \equiv \frac{\alpha_{,v}}{\alpha},\quad f \equiv r_{,u},\quad g \equiv r_{,v},\quad w \equiv s_{,u},\quad z \equiv s_{,v}.
\end{eqnarray}

The Einstein tensors are then given as follows:
\begin{eqnarray}
G_{uu} &=& -\frac{1}{r} \left(f_{,u}-2fh \right),\\
G_{uv} &=& \frac{f_{,v}}{r},\\
G_{vv} &=& -\frac{1}{r} \left(g_{,v}-2gd \right),\\
G_{\varphi\varphi} &=& -\frac{4r^{2}}{\alpha^{2}} d_{,u}.
\end{eqnarray}

Via the gauge symmetry of the Maxwell field, we can choose the Maxwell field $A_{\mu} = (a, 0, 0)$ by introducing the function $a(u,v)$. Thus, we can obtain the energy-momentum tensors of the classical part with the cosmological constant term by:
\begin{eqnarray}
T^{\mathrm{C}}_{uu} - \frac{\Lambda}{8 \pi} g_{uu} &=& \frac{1}{4\pi} \left[ w\overline{w} + iea(\overline{w}s-w\overline{s}) +e^{2}a^{2}s\overline{s} \right],\\
T^{\mathrm{C}}_{uv} - \frac{\Lambda}{8 \pi} g_{uv} &=& \frac{{(a_{,v})}^{2}}{4 \pi \alpha^{2}} + \frac{\alpha^{2}}{16\pi}\Lambda,\\
T^{\mathrm{C}}_{vv} - \frac{\Lambda}{8 \pi} g_{vv} &=& \frac{1}{4\pi} z\overline{z},\\
T^{\mathrm{C}}_{\varphi\varphi} - \frac{\Lambda}{8 \pi} g_{\varphi\varphi} &=& \frac{r^{2}}{4 \pi \alpha^{2}} \left[ (w\overline{z}+z\overline{w}) + iea(\overline{z}s-z\overline{s}) + \frac{2 {(a_{,v})}^{2}}{\alpha^{2}} \right] - \frac{r^{2}}{8\pi}\Lambda.
\end{eqnarray}

Note that $F^{\mu}\,_{\nu;\mu} = 4\pi J_{\nu}$, where $J_{\nu}$ is asymptotically interpreted by the electric current. If we define $q(u,v) \equiv r a_{,v}/\alpha^{2}$ as the charge function, then $J_{\nu} = (1/2\pi r)(-q_{,u},q_{,v},0)$ and hence corresponds to the concept of the charge.

There are several discussions pertaining to re-normalized energy-momentum tensors for three dimensions in the literature \cite{Steif:1993zv}. As expected for two- or four-dimensional cases \cite{cc}\cite{Poisson:1997my}, the re-normalized energy-momentum tensors diverge along the inner horizon for the static metric. However, we know that the internal structures cannot be described by the static metric for fully dynamical cases. Due to dynamical situations and spherical symmetry, therefore, the following represents the best choice for the re-normalized energy-momentum tensor: use the $(1+1)$-dimensional results \cite{Davies:1976ei}\cite{Birrell:1982ix} divided by $2\pi r$ and apply the $S$-wave approximation by
\begin{eqnarray}
\langle \hat{T}^{\mathrm{H}}_{uu} \rangle &=& \frac{P}{2\pi r}\left(h_{,u}-h^{2}\right)\label{eq:semi1},
 \\
\langle \hat{T}^{\mathrm{H}}_{uv} \rangle = \langle \hat{T}^{\mathrm{H}}_{vu} \rangle &=& -\frac{P}{2\pi r}d_{,u}\label{eq:semi2},
 \\
\langle \hat{T}^{\mathrm{H}}_{vv} \rangle &=& \frac{P}{2\pi r}\left(d_{,v}-d^{2}\right)\label{eq:semi3},
\end{eqnarray}
where $P \equiv \mathcal{N} \hbar / 12\pi$ and $\mathcal{N}$ is the number of massless scalar fields.

After solving these coupled equations, we can write all equations:
\begin{eqnarray}
\label{eq:E1}f_{,u} &=& 2fh - 8 \pi r \left(T^{\mathrm{C}}_{uu} + \langle \hat{T}^{\mathrm{H}}_{uu} \rangle \right),\\
\label{eq:E2}g_{,v} &=& 2gd - 8 \pi r \left(T^{\mathrm{C}}_{vv} + \langle \hat{T}^{\mathrm{H}}_{vv} \rangle \right),\\
\label{eq:E3}g_{,u}=f_{,v} &=& 8 \pi r \left( T^{\mathrm{C}}_{uv} - \frac{\Lambda}{8 \pi} g_{uv} \right) + 8 \pi P \frac{\alpha^{2}}{r^{2}} \left( T^{\mathrm{C}}_{\varphi \varphi} - \frac{\Lambda}{8 \pi} g_{\varphi \varphi} \right),\\
\label{eq:E4}d_{,u}=h_{,v} &=& - \frac{2 \pi \alpha^{2}}{r^{2}} \left( T^{\mathrm{C}}_{\varphi \varphi}- \frac{\Lambda}{8 \pi} g_{\varphi \varphi} \right),
\end{eqnarray}
including the field equations
\begin{eqnarray}
\label{eq:M1}a_{,v} &=& \frac{\alpha ^{2} q}{r}, \\
\label{eq:M2}q_{,v} &=& -\frac{ier}{4} (\overline{s}z-s\overline{z}), \\
\label{eq:s}z_{,u} = w_{,v} &=& - \frac{fz}{2r} - \frac{gw}{2r} - ieaz - \frac{ieags}{2r} - \frac{ie}{2r}\alpha^{2}qs.
\end{eqnarray}

At this point, the equations of $\alpha_{,uv}$, $r_{,uv}$, and $s_{,uv}$ parts can be represented by first-order differential equations.
We can then implement the same integration scheme used in previous papers \cite{Hong:2008mw}\cite{Hong:2008mw_2}.
We used the second-order Runge-Kutta method \cite{nr}; however, due to the differentials in Equations~(\ref{eq:E1}) and (\ref{eq:E2}), our results will converge to the first order if $P>0$. Tests of the convergence are provided in Appendix~B.

\subsection{Definition of mass function}

One important question when interpreting our results is whether or not these black hole solutions are BTZ solutions. Of course, due to the Birkhoff theorem for three-dimensional cases, we can be sure that our solutions will be BTZ (for further discussions, see Appendix~A). Hence, there will be a natural concept for the mass function.

First, we can define the general spherical symmetric metric in $(D+1)$ dimensions in the static limit by the following coordinates:
\begin{eqnarray}
ds^{2} &=& - N^{2} dt^{2} + \frac{dr^{2}}{N^{2}} + r^{2} d\Omega_{D-1}^{2}\\
&=& - N^{2} dU dV + r^{2}(U, V) d\Omega_{D-1}^{2},
\end{eqnarray}
where
\begin{eqnarray}
dU &=& dt - \frac{dr}{N^{2}},\\
dV &=& dt + \frac{dr}{N^{2}}.
\end{eqnarray}

In many cases, function $N$ vanishes near the horizon; hence, it will be convenient to redefine the coordinate in the following form:
\begin{eqnarray}
ds^{2} = - \alpha^{2}(u,v) du dv + r^{2}(u, v) d\Omega_{D-1}^{2},
\end{eqnarray}
where the coordinate transformation is
\begin{eqnarray}
dr &=& r_{,u}du + r_{,v}dv,\\
dt &=& \frac{\alpha^{2}}{4} \left( -\frac{dv}{r_{,u}} + \frac{du}{r_{,v}} \right).
\end{eqnarray}
Thus, we can show
\begin{eqnarray}
N^{2} = - \frac{4r_{,u}r_{,v}}{\alpha^{2}}.
\end{eqnarray}
Note that $r_{,u}r_{,v}$ is negative outside of the black hole. Therefore, the signature is fine. Moreover, we can easily see that the apparent horizons $r_{,v}=0$ or $r_{,u}=0$ make $N=0$, although $\alpha$ is not zero.

Thus, in the four dimensions, we can define the mass function \cite{Waugh:1986jh} by
\begin{eqnarray}
m_{(4)}(u,v) = \frac{r}{2} \left( 1 + \frac{4r_{,u}r_{,v}}{\alpha^{2}} + \frac{q^{2}}{r^{2}} - \frac{\Lambda}{3}r^{2} \right).
\end{eqnarray}
Also, for three dimensions, we can similarly define the mass aspect by
\begin{eqnarray}
m_{(3)}(u,v) = \frac{r^{2}}{l^{2}} - q^{2} \ln \frac{r^{2}}{l^{2}} + \frac{4r_{,u}r_{,v}}{\alpha^{2}}.
\end{eqnarray}
This definition is consistent with the findings in the literature \cite{Alcubierre:1999ex}.

\begin{figure}
\begin{center}
\includegraphics[scale=0.75]{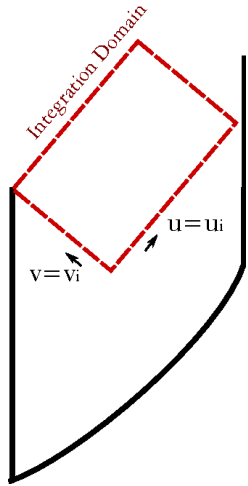}
\caption{\label{fig:domain}Integration domain of our simulations. We assign initial conditions along the in-going and out-going null surfaces along the arrows.}
\end{center}
\end{figure}
\begin{table}[b]
\begin{center}
\begin{tabular}{c|c|c|c}
\hline
& \;\;Fixed by hand\;\; & \;\;Fixed by constraints\;\; & \;\;Fixed by evolution equations\;\;\\
\hline \hline
$v=v_{\mathrm{i}}$ & $s=0, f, r$ & $w=h=a=q=0$, $\alpha=1$ & $d, g, z$\\
\hline
$u=u_{\mathrm{i}}$ & $s, g, r$ & $z=s_{,v}$, $d=2rz\bar{z}$, $\alpha \leftarrow d=\alpha_{,v}/\alpha$ & $h, f, w, q, a$\\
\hline
\end{tabular}
\caption{\label{table:conditions}Summary of the assignments of initial conditions for $v=v_{\mathrm{i}}$ (the in-going initial surface) and $u=u_{\mathrm{i}}$ (the out-going initial surface).}
\end{center}
\end{table}

\subsection{\label{sec:ini}Initial conditions and free parameters}

We need initial conditions for all functions ($\alpha, h, d, r, f, g, s, w, z, a, q$) on the initial $u=u_{\mathrm{i}}$ and $v=v_{\mathrm{i}}$ surfaces, where we set $u_{\mathrm{i}}=v_{\mathrm{i}}=0$ (Figure~\ref{fig:domain}).

We have gauge freedom to choose the initial $r$ function. Although all constant $u$ and $v$ lines are null, there remains freedom to choose the distances between these null lines. Here, we choose $r(0,0)=r_{0}$, $f(u,0)=r_{u0}$, and $g(0,v)=r_{v0}$, where $r_{u0}<0$ and $r_{v0}>0$ such that the radial function for an in-going observer decreases and that for an out-going observer increases.

We use a shell-shaped scalar field. Therefore, its interior is not affected by the shell. First, it is convenient to choose $r_{u0}=-1/2$ and $r_{v0}=1/2$; the mass function on $u_{\mathrm{i}}=v_{\mathrm{i}}=0$ then becomes $m_{(3)}(u_{\mathrm{i}},v_{\mathrm{i}}) = r_{0}^{2}/l^{2}-1/\alpha(u_{\mathrm{i}},v_{\mathrm{i}})^{2}$, as initially there is no charge. The pure anti de Sitter limit is $m_{(3)}=-1$, and hence as long as $m_{(3)}>-1$, the choice of $\alpha(u_{\mathrm{i}},v_{\mathrm{i}})$ is physically allowed, as one can always assume that there is a certain core of mass at the center. In this paper, we choose $\alpha(u,0)=1$ for convenience. Also, $s(u,0)=w(u,0)=h(u,0)=a(u,0)=q(u,0)=0$ hold.

We need more information to determine $d, g$, and $z$ on the $v=0$ surface. We obtain $d$ from Equation~(\ref{eq:E4}), $g$ from Equation~(\ref{eq:E3}), and $z$ from Equation~(\ref{eq:s}) (Table~\ref{table:conditions}).

We can choose an arbitrary function for $s(0,v)$ to induce a collapsing pulse. In this paper, we use
\begin{eqnarray} \label{s_initial}
s(u_{\mathrm{i}},v)= A \sin^{2} \left( \pi \frac{v-v_{\mathrm{i}}}{v_{\mathrm{f}}-v_{\mathrm{i}}} \right) \exp \left( 2 \pi i \frac{v-v_{\mathrm{i}}}{v_{\mathrm{f}}-v_{\mathrm{i}}} \right)
\end{eqnarray}
for $v_{\mathrm{i}}\leq v \leq v_{\mathrm{f}}$ and otherwise $s(u_{\mathrm{i}},v)=0$, where $u_{\mathrm{i}}=0$, $v_{\mathrm{i}}=0$, and $v_{\mathrm{f}}=20$ denotes the end of the pulse in the initial surface.
We then obtain $z(0,v)=s(0,v)_{,v}$.
This implements one pulse of energy ($T_{vv} \sim z^{2}$) along the out-going null direction by the continuous function $z(0,v)$.

Also, from Equation~(\ref{eq:E2}), we can use $d = 2 r z \bar{z}$ on the $u=0$ surface, given the assumption that there is no Hawking effect on the initial surface. We then obtain $d(0,v)$. By integrating $d$ along $v$, we have $\alpha(0,v)$.

We need more information for $h, f, q, a,$ and $w$ on the $u=0$ surface. We obtain $h$ from Equation~(\ref{eq:E4}), $f$ from Equation~(\ref{eq:E3}), $a$ from Equation~(\ref{eq:M1}), $q$ from Equation~(\ref{eq:M2}), and $w$ from Equation~(\ref{eq:s}). This finishes the assignments of the initial conditions (Table~\ref{table:conditions}).

We choose $r_{0}=10$, $\Lambda/8\pi = -0.0001$, and $A=0.2$, leaving the two free parameters $(e, P)$, where $e$ is the gauge coupling and $P$ determines the strength of the semi-classical effects.

\subsection{\label{sec:dim}Dimensional analysis}

For four-dimensional cases, if we define $c=G_{4}=1$, then the Planck units (Planck length $l_{\mathrm{Pl,4}}$ and Planck mass $m_{\mathrm{Pl,4}}$) become
\begin{eqnarray}
l_{\mathrm{Pl,4}} = m_{\mathrm{Pl,4}} = \sqrt{\hbar}.
\end{eqnarray}
Because $P \propto \mathcal{N} \hbar$ should be fixed by a simulation parameter, as $\mathcal{N}$ becomes larger and larger, $\hbar$ becomes smaller and smaller. For example, for four-dimensional simulations, if we made a black hole with size $r_{+} = 10$, we fixed $P = 1$, and we assume $\mathcal{N} = 100$, then the physical size of the black hole in Planck units is $r_{+} = 100\, l_{\mathrm{Pl,4}}$. By the same argument, the `bare' simulation results with dimension length $L$, mass $M$, or curvature $R$ (e.g., the Ricci scalar) are re-scaled in the Planck units by
\begin{eqnarray}
L &=& \sqrt{\mathcal{N}} L l_{\mathrm{Pl,4}}, \\
M &=& \sqrt{\mathcal{N}} M m_{\mathrm{Pl,4}}, \\
R &=& \frac{R}{\mathcal{N}} l^{-2}_{\mathrm{Pl,4}},
\end{eqnarray}
where $\mathcal{N} \hbar = \mathcal{N} l^{2}_{\mathrm{Pl,4}} = \mathcal{N} m^{2}_{\mathrm{Pl,4}} = 1$ \cite{Yeom:2009zp}.

For three-dimensional cases, the dimensional analysis is different. If $c = G_{3} =1$, then
\begin{eqnarray}
l_{\mathrm{Pl,3}} = \hbar
\end{eqnarray}
and $m_{\mathrm{Pl,3}}$ is dimensionless.
Again, $P$ can be fixed as a constant (e.g., $1$), and then any bare quantities with dimensions length $L$, mass $M$, or curvature $R$ are re-scaled in Planck units by
\begin{eqnarray}
L &=& \mathcal{N} L l_{\mathrm{Pl,3}}, \\
M &=& M m_{\mathrm{Pl,3}}, \\
R &=& \frac{R}{\mathcal{N}^{2}} l^{-2}_{\mathrm{Pl,3}},
\end{eqnarray}
where $\mathcal{N} \hbar = \mathcal{N} l_{\mathrm{Pl,3}} = 1$.

One interesting difference between the three-dimensional and four-dimensional results is the large-$\mathcal{N}$ cutoff. According to Dvali \cite{Dvali:2007hz}, when there is a large number $\mathcal{N}$ of particle species that contribute in the mode of Hawking radiation, the cutoff scale should be stretched by
\begin{eqnarray}
l_{\mathrm{cutoff,4}} = \sqrt{\mathcal{N}} l_{\mathrm{Pl,4}}.
\end{eqnarray}
Dvali supported this idea using arguments that included black hole complementarity \cite{Susskind:1993if}: if a black hole is smaller than $l_{\mathrm{cutoff,4}}$, then the evaporation is too fast and one cannot regard it as a semi-classical black hole. In such a case, it can violate the principle of black hole complementarity.\footnote{Also, see further discussions on black hole complementarity in \cite{Yeom:2008qw}.}

In the double-null formalism, for four-dimensional cases the Dvali cutoff is naturally implemented via the $(1-P/r^{2})^{-1}$ term in an equation \cite{Hong:2008mw}. Thus, $\sqrt{P} \sim \sqrt{N} l_{\mathrm{Pl,4}}$ was the natural cutoff scale of the simulation. Therefore, any results of the double-null simulation was consistent with Dvali's argument.
However, in the three-dimensional case, there is no singular term despite the fact that we can choose a large $P$ parameter. It is well known that a black hole is in the thermal equilibrium and does not totally evaporate in three-dimensions \cite{Reznik:1991qj}; hence, we do not have to worry about the information loss problem. Therefore, there is no essential reason to consider a large-$\mathcal{N}$ cutoff. This thermodynamic intuition is naturally implemented in our double-null formalism.

On the other hand, the correction term from the re-normalized energy-momentum tensor for four dimensions becomes
\begin{eqnarray}
\langle {\hat{T}^{\mathrm{H}}_{\mu\nu}} \rangle \sim \frac{\mathcal{N} \hbar}{r^{2}} (\mathrm{dimensionless\; terms}) + \mathcal{O}(\hbar^{2}) + ...,
\end{eqnarray}
while for three-dimensional cases,
\begin{eqnarray}
\langle {\hat{T}^{\mathrm{H}}_{\mu\nu}} \rangle \sim \frac{\mathcal{N} \hbar}{r} (\mathrm{dimensionless\; terms}) + \mathcal{O}(\hbar^{2}) + ...
\end{eqnarray}
occurs. Therefore, $r = \sqrt{P}$ in four dimensions and $r = P$ for three dimensions are a type of turning point during which the contribution of the re-normalized energy-momentum tensor becomes dominant over the other terms. If this intuition is correct, the inside of the $r=P$ region for three-dimensional cases is dominated by negative energy via semi-classical effects rather than classical matters. This may imply that $r \lesssim P$ is a physically special region that changes the classical properties, although it is not a definite cutoff as in four-dimensional cases.

\section{\label{sec:dyn}Dynamical formation and evolution of three-dimensional black holes}

\subsection{Gravitational collapses without Hawking radiation}

\begin{figure}
\begin{center}
\includegraphics[scale=0.3]{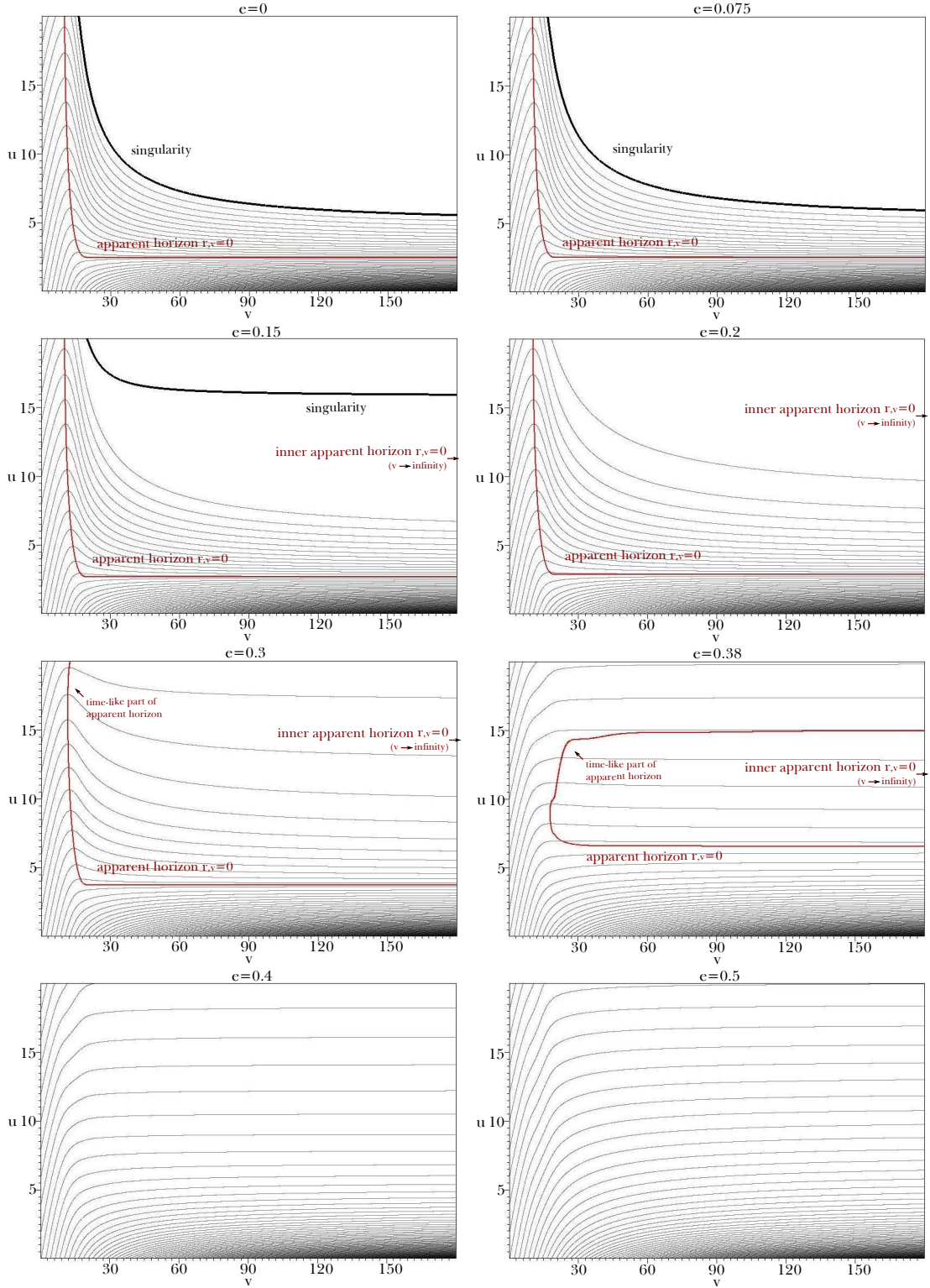}
\caption{\label{fig:P=0}Numerical results for $P=0$ (without Hawking radiation) while $e$ varies.}
\end{center}
\end{figure}

We study the gravitational collapse of a charged matter shell and the formation of black holes without Hawking radiation. We varied the gauge coupling parameter $e$ to change the amount of charge of the matter shell. Figure~\ref{fig:P=0} is the result of the numerical simulations.

\subsubsection{\label{sec:cau}Causal structures}

\begin{figure}
\begin{center}
\includegraphics[scale=0.6]{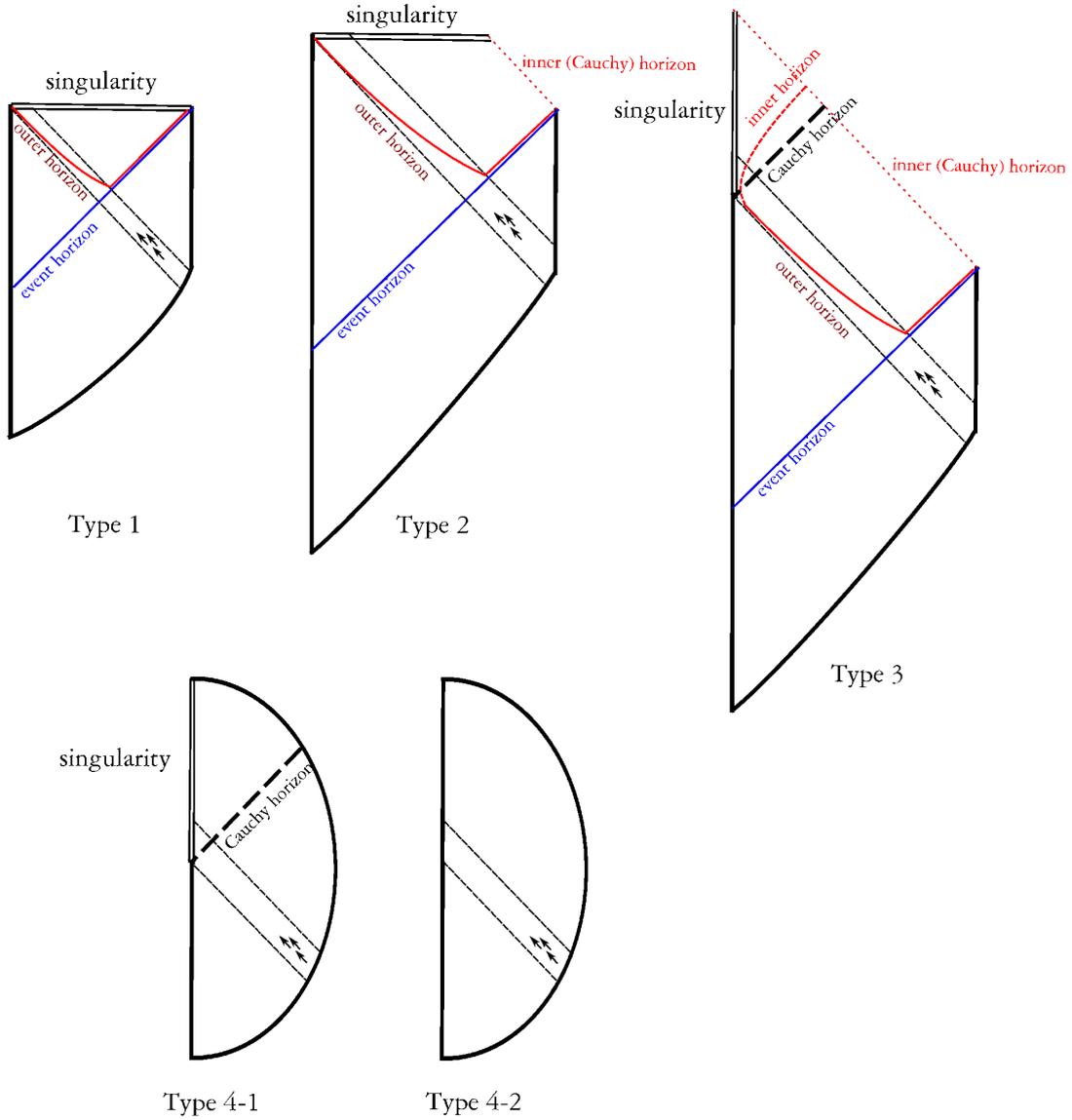}
\caption{\label{fig:P0_types}Causal structures without Hawking radiation. Type~$1$: $Q=0$ case. Type~$2$: if the charge is sufficiently small, there can be a space-like singularity as well as a null inner Cauchy horizon in the $v \rightarrow \infty$ limit. Type~$3$: if the charge is sufficiently large, one cannot see a space-like singularity. The singularity is then beyond the inner horizon and it must be time-like; therefore, it is a reasonable to guess that there is a time-like inner horizon beyond the Cauchy horizon. Type~$4$-$1$ and Type~$4$-$2$: if the charge is excessive, the horizon then disappears. In this case, there may be a time-like singularity beyond the Cauchy horizon. Or, the charged matter may be repulsed via the strong electric field and fail to form a singularity.}
\end{center}
\end{figure}

For the $e=0$ limit, there is no gauge coupling and the black hole is neutral. There is a space-like singularity and an apparent horizon. The apparent horizon grows in the space-like direction during the matter collapse ($v<v_{\mathrm{f}}=20$) and approaches a null direction (Type~$1$ in Figure~\ref{fig:P0_types}).

Also, if gauge coupling is sufficiently small so that the amount of charge is negligible ($e=0.075$), the causal structure is similar to the neutral case. However, the location of the singularity changes slightly. As we increase the parameter $e$, the location of the space-like singularity drastically changes ($e=0.15$).

If we increase $e$ more, we cannot see the space-like singularity ($e=0.2$). The result of $e=0.2$ itself does not offer information about the existence or absence of the space-like singularity. However, in the $e=0.3$ and $e=0.38$ cases, we can see a time-like piece of the apparent horizon during the matter collapse. Although the apparent horizon is time-like, it does not violate the null energy condition, as an out-going observer sees a decreasing $r$. Rather, the time-like part should be identified with a type of inner horizon of the black hole; the outer and inner horizons are separated during gravitational collapses. Thus, although we cannot see beyond the $u=20$ surface, we can guess that the inner horizon should be time-like. Also, we believe that there should be a singularity around the center via the singularity theorem, and that it may be inside of the inner horizon. If this expectation is reasonable, then the singularity should be time-like.

\begin{figure}
\begin{center}
\includegraphics[scale=0.75]{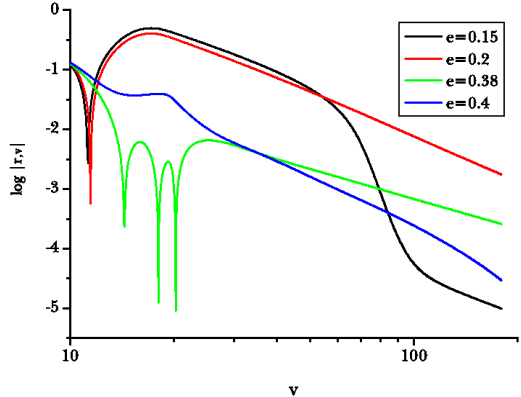}
\caption{\label{fig:G_behavior}$\log |r_{,v}|$ versus $\log v$ along $u=10$ for $e=0.15$, $0.2$, $0.38$, and $0.4$. The gradients for a large $v$ limit are $-2.2951$, $-2.4932$, $-1.6447$, and $-3.7969$, respectively.}
\end{center}
\end{figure}

Now let us comment on the behavior of $r_{,v}$ with the large $v$ limit. In Figure~\ref{fig:G_behavior}, we plot the behaviors of $\log |r_{,v}|$ as a function of $\log v$ along $u=10$. Note that for $e=0.15$, $0.2$, and $0.38$, there are horizons and the $u=10$ surface will be included by the horizon, while the $u=10$ line for $e=0.4$ is not trapped by a horizon. The plot shows that $|r_{,v}|$ converges $0$ for a large $v$ limit as a function of $v^{-\beta}$, where $\beta > 1$ is approximately $2.3$, $2.5$, $1.6$, and $3.8$ respectively for the four cases. Note that if $\beta$ is larger than $1$, then
\begin{eqnarray}
r = \int r_{,v} dv \sim v^{-\beta+1} + \mathrm{const}
\end{eqnarray}
will converge for $v \rightarrow \infty$ limit. This implies that, in the $v \rightarrow \infty$ limit, the inside of the horizon becomes a null $r_{,v}=0$ horizon with a finite $r$. This behavior arises even for the $e=0.4$ case, as it is in the anti de Sitter background. In this case, a large $v$ limit should be regarded as the time-like boundary of the anti de Sitter space, implying that an outgoing observer cannot experience an increase of the area.

Therefore, in terms of the inside of the outer apparent horizon, as in the four-dimensional cases, given the large $v$ limit, we can interpret these as an inner Cauchy horizon ($r_{,v}\rightarrow 0$). In this case, the inner horizon is null. Therefore, we can determine the causal structure for a small charge case (Type~$2$ in Figure~\ref{fig:P0_types}) and for a large charge case (Type~$3$ in Figure~\ref{fig:P0_types}).

If we increase $e$ until it is excessively large, we cannot see apparent horizons and the black hole disappears. This phenomenon appears odd. For four-dimensional cases, we can make an asymptotic matter shell with $M<Q$, but a black hole always forms for any field combinations \cite{Hong:2008mw}. This means that, for four-dimensional cases, the excessive charge is scattered and repulsed via electromagnetic forces (by purely classical effects). This makes it impossible to see a naked singularity ($Q>M$ of Reissner-Nordstr\"{o}m solutions) and conserves weak cosmic censorship. However, for three-dimensional cases, we cannot not see a horizon in this case. To interpret this, there are two possibilities: a naked singularity (Type~$4$-$1$ in Figure~\ref{fig:P0_types}) or a regular center (Type~$4$-$2$ in Figure~\ref{fig:P0_types}). In principle, it is impossible to decide for certain, but we can offer a reasonable guess as to what the correct answer is, as shown in Section~\ref{sec:cri}.

\subsubsection{\label{sec:cri}Black hole formation near extreme limit}

\begin{figure}
\begin{center}
\includegraphics[scale=0.3]{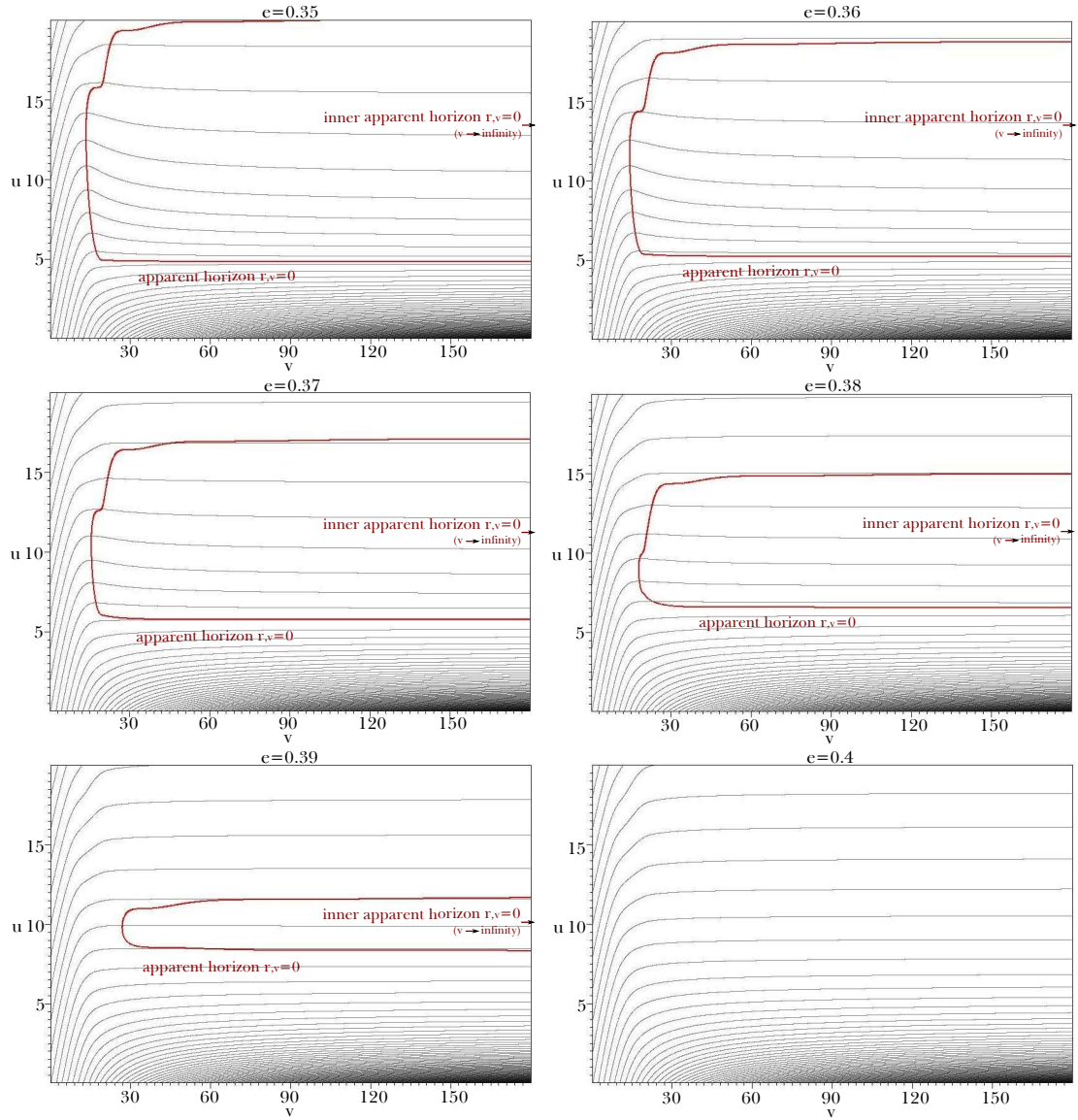}
\caption{\label{fig:P=0_critical}Numerical results for $P=0$ while varying from $e=3.5$ to $e=4.0$.}
\end{center}
\end{figure}

We observe the details of causal structures between $e=0.35$ and $e=0.4$, where a type of critical behavior occurs (Figure~\ref{fig:P=0_critical}). As we increase the charge, two horizons (time-like inner and space-like outer apparent horizons) approach. Additionally, the advanced time $v$ when the apparent horizon appears seems to increase so that the large charge postpones the formation of the apparent horizon. The location of the apparent horizon decreases, which implies that the effective mass of a black hole decreases as the charge increases.

We know that there will not be horizons when $e=0.4$ or $e=0.5$ between $v=0$ and $v=180$, even beyond the $u=20$ line (we know that the horizons disappeared as the charge increases, as shown in Figure~\ref{fig:P=0_critical}).
Therefore, if there is a singularity for such cases, it should imply that it is a type of naked singularity. If this is true, then it is fair to say that we cannot see beyond the $u=20$ line in principle. However, reasonable assumptions can be made given some circumstantial evidence and extrapolations.

\begin{figure}
\begin{center}
\includegraphics[scale=0.75]{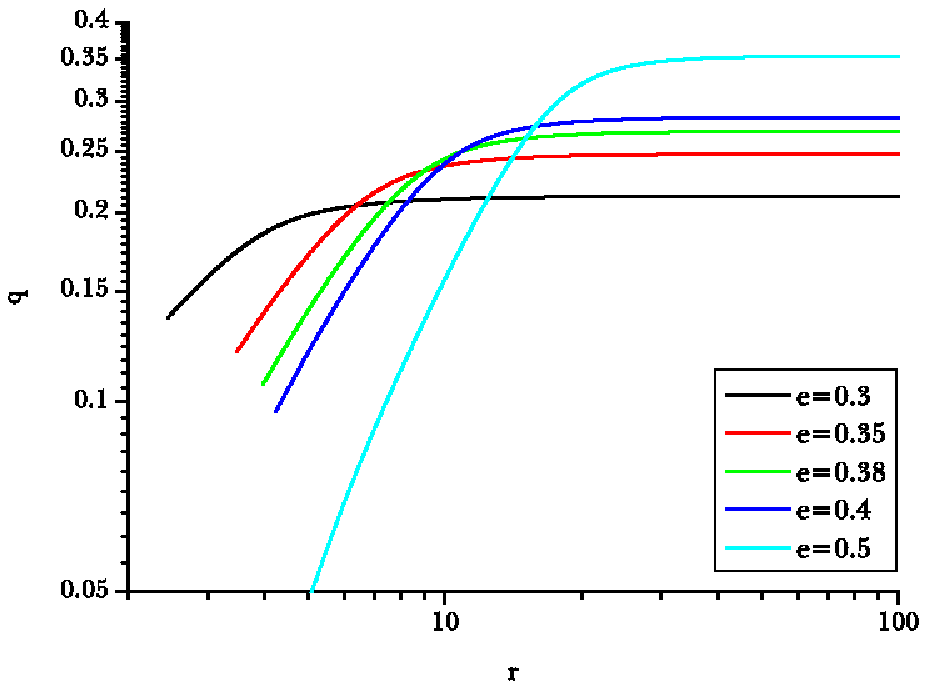}
\caption{\label{fig:q_r}Charge $q(r)$ at $v=180$ for each $e$.}
\end{center}
\end{figure}
\begin{figure}
\begin{center}
\includegraphics[scale=0.75]{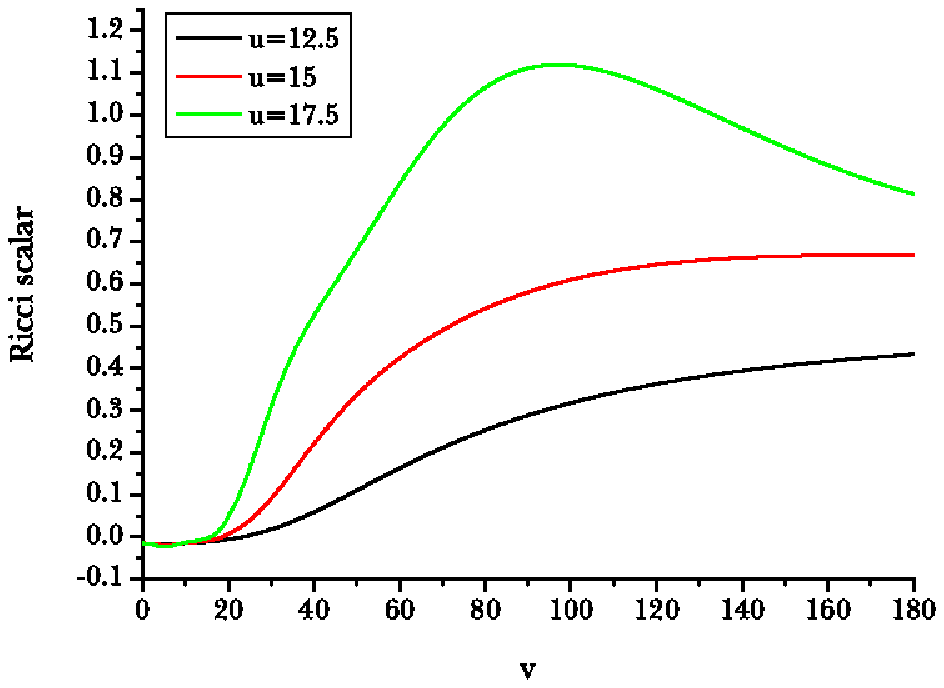}
\caption{\label{fig:Ricci_P0}Ricci scalar for $u=12.5$, $15$, and $17.5$, for the $e=0.2$ case.}
\end{center}
\end{figure}

We plot charge $q$ as a function of $r$ at the advanced time $v=180$ in Figure~\ref{fig:q_r}. As the charge increases, the charge of outside (large $r$) increases while the charge of inside (small $r$) decreases. This implies that our field combination asymptotically induces a large charge, as we use a large coupling $e$. However, as the charge increases, mutual repulsion is noted. Hence, a collapse near the center is not probable, and the center has less charge as $e$ increases.

Note that this diagram is a log-log plot and that appears to be linear for a large $r$ and a small $r$. Of course, we are interested in the charge distribution near the center, and hence, an estimation of the gradient of the $\log q$-$\log r$ plot near the center, i.e., the exponent of $q=A r^{B}$ near the center, will serve as a viable means of extrapolation beyond the $u=20$ surface.

For each case, we calculated the exponents in Table~\ref{table:fitting}. Here, the energy density for a constant $r$ observer will be
\begin{eqnarray}
\rho = -T^{t}_{\;t} = -\frac{1}{r_{,u}r_{,v}\alpha^{2}} \left( r_{,v}^{2} T_{uu} - 2 r_{,u}r_{,v} T_{uv} + r_{,u}^{2} T_{vv} \right),
\end{eqnarray}
where the dominant term in $r \rightarrow 0$ limit will be proportional to
\begin{eqnarray}
\rho \propto \left( \frac{q}{r} \right)^{2},
\end{eqnarray}
where it is proportional to the electric energy density.
Thus, the electric energy density is proportional to
\begin{eqnarray}
\rho(r) \propto \left( \frac{q(r)}{r} \right)^{2} \sim r^{2(B-1)}.
\end{eqnarray}
Therefore, roughly speaking, if $B$ is less than $1$, it will be evidence that the energy density $\rho$ diverges around the center, showing that there is a central singularity.

\begin{table}
\begin{center}
\begin{tabular}{c|c|c}
\hline
\;\;\;\;\;\;\;\; $e$ \;\;\;\;\;\;\;\; & \;\;\;\;\;\;\;\; $A$ \;\;\;\;\;\;\;\; & \;\;\;\;\;\;\;\; $B$ \;\;\;\;\;\;\;\; \\
\hline \hline
$0.3$ & $0.0694$ & $0.7534$ \\
\hline
$0.35$ & $0.0332$ & $1.0327$ \\
\hline
$0.38$ & $0.0203$ & $1.2035$ \\
\hline
$0.4$ & $0.0144$ & $1.3195$ \\
\hline
$0.5$ & $0.0019$ & $2.0176$ \\
\hline
\end{tabular}
\caption{\label{table:fitting}Numerical fitting for $q(r)$ to the functional form $q(r) = A r^{B}$ for each value of $e$. We fitted from $u=19$ to $u=20$ to determine the behavior around the center.}
\end{center}
\end{table}

Therefore, we may conclude that, when $e=0.3$, there is surely a central singularity while for the other cases this is not evident. Of course, via the singularity theorem, if there is an apparent horizon, there will be a singularity anyway. For the $e=0.4$ and $0.5$ cases, however, there is no horizon. In these cases, \textit{these exponents are circumstantial evidence that there may not be a singularity around the center.} (This conclusion is consistent with the result of \cite{Alcubierre:1999ex}.) Of course, if the distribution of the charge drastically changes beyond $u=20$, the central structure will be modified and our extrapolation will be meaningless. Therefore, our prospect depends, in all respects, on circumstantial evidence and not on real evidence.

\begin{figure}
\begin{center}
\includegraphics[scale=0.75]{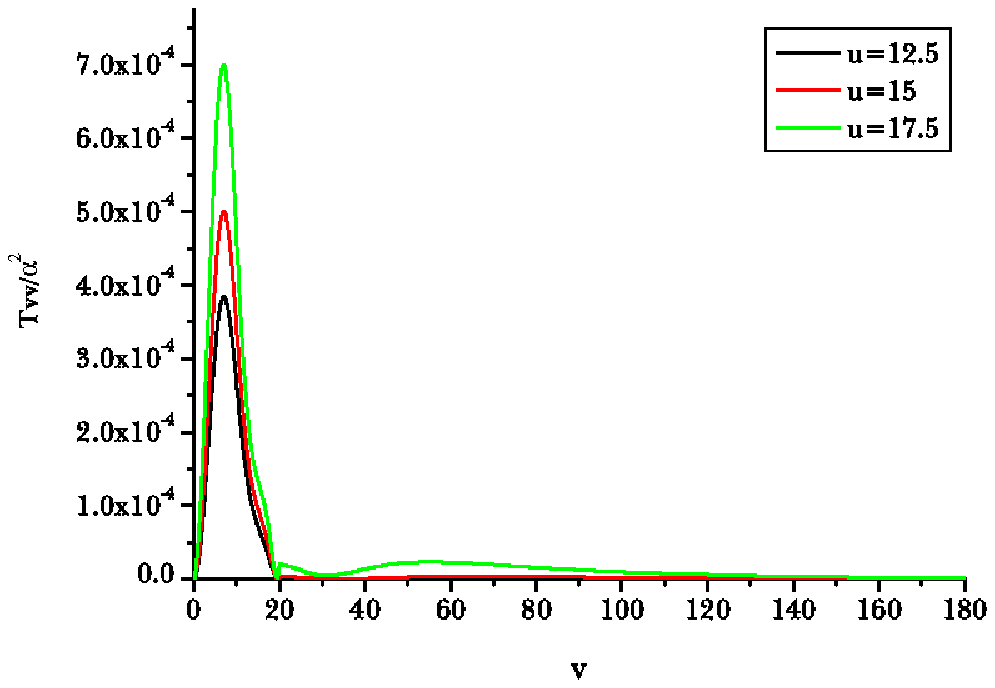}
\includegraphics[scale=0.75]{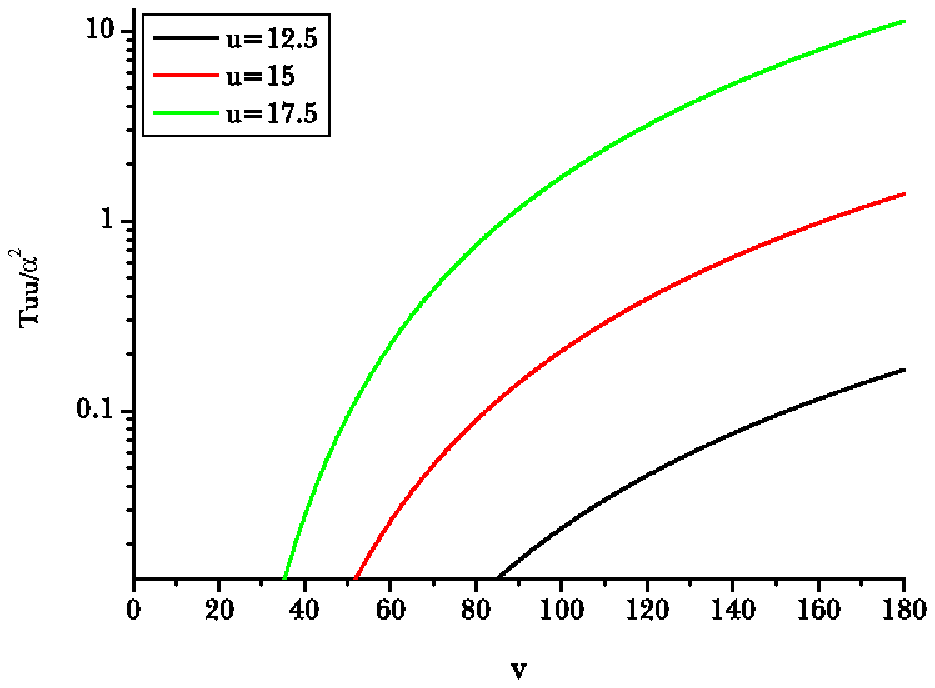}
\includegraphics[scale=0.75]{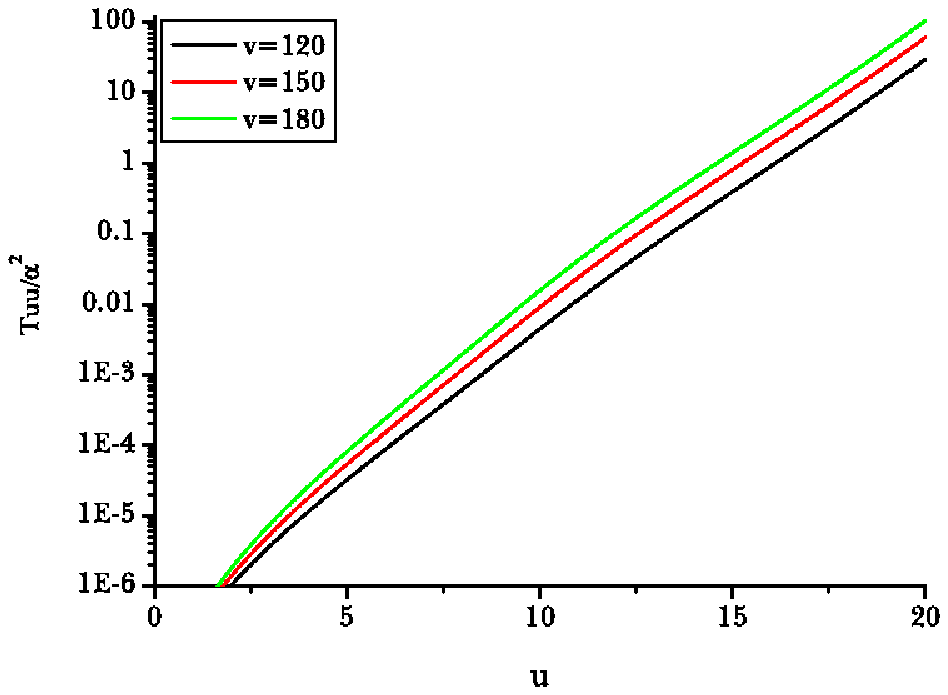}
\caption{\label{fig:Tuu_P0_2}Top and middle: $T_{vv}/\alpha^{2}$ and $T_{uu}/\alpha^{2}$ for $u=12.5$, $15$, and $17.5$, for the $e=0.2$ case. Bottom: $T_{uu}/\alpha^{2}$ for $v=120$, $150$, and $180$, for the $e=0.2$ case.}
\end{center}
\end{figure}

\subsubsection{\label{sec:mas}Mass inflation}

There are several studies in the literature that closely investigated the mass inflation in three-dimensional black holes \cite{Cai:1995ib}. One of the interesting predictions was that there is a curvature singularity for a certain curvature component, whereas some of the other curvature components do not diverge as $v$ increases.

We observed the Ricci scalar $R$ (Figure~\ref{fig:Ricci_P0}), where
\begin{eqnarray}
R = \frac{8}{\alpha^{2}} \left( \frac{r_{,uv}}{r} + \frac{\alpha \alpha_{,uv}-\alpha_{,u}\alpha_{,v}}{\alpha^{2}} \right),
\end{eqnarray}
$T_{\mu\nu}l^{\mu}l^{\nu} \propto T_{vv}/\alpha^{2}$ for an out-going null observer $l^{\mu}$ (top in Figure~\ref{fig:Tuu_P0_2}), $T_{\mu\nu}n^{\mu}n^{\nu} \propto T_{uu}/\alpha^{2}$ for an in-going null observer $n^{\mu}$, for some out-going (middle in Figure~\ref{fig:Tuu_P0_2}) and in-going null slices (bottom in Figure~\ref{fig:Tuu_P0_2}) ($u=12.5$, $15$, and $17.5$; $v=120$, $150$, and $180$). Note that we choose ($[u,v,\varphi]$)
\begin{eqnarray}
l^{\mu}&=&\frac{\sqrt{2}}{\alpha}(0,1,0),\\
n^{\mu}&=&\frac{\sqrt{2}}{\alpha}(1,0,0),
\end{eqnarray}
so that $l^{\mu}l_{\mu}=n^{\mu}n_{\mu}=0$ and $l^{\mu}n_{\mu}=-1$. Thus, any time-like geodesic $t^{\mu}$ can be decomposed by
\begin{eqnarray}
t^{\mu}= a l^{\mu} + b n^{\mu}
\end{eqnarray}
with constraint $ab=1/2$. Hence, any time-like observer will see local energy density by
\begin{eqnarray}
T_{\mu\nu} t^{\mu}t^{\nu}=\frac{2}{\alpha^{2}} \left( b^{2} T_{uu} + a^{2} T_{vv} \right) + \frac{q^{2}}{2\pi r^{2}}.
\end{eqnarray}
Therefore, if there is a signature for a case in which $T_{uu}/\alpha^{2}$ or $T_{vv}/\alpha^{2}$ increases exponentially, then it is the evidence of mass inflation.

The results are interesting: there is no evidence that the Ricci scalar or $T_{vv}/\alpha^{2}$ diverge, while $T_{uu}/\alpha^{2}$ shows exponential growth. Thus, this provides clear evidence that there is mass inflation and that a curvature singularity forms around the inner Cauchy horizon at the $v \rightarrow \infty$ limit.
However, this behavior of curvature functions is very different from that noted in four-dimensional cases. For four-dimensional cases, almost all related functions (e.g., Ricci scalar, Kretchmann scalar, Misner-Sharp mass function) diverge exponentially, while the Ricci scalar and $T_{vv}/\alpha^{2}$ components do not diverge in three dimensions. Therefore, this is a typical property of three-dimensional charged black holes.

\subsection{Semi-classical back-reactions by Hawking radiation}

In this section, now we turn to the semi-classical effects and see dynamical back-reactions for the inside structures.

\subsubsection{\label{sec:res}Results}

\begin{figure}
\begin{center}
\includegraphics[scale=0.3]{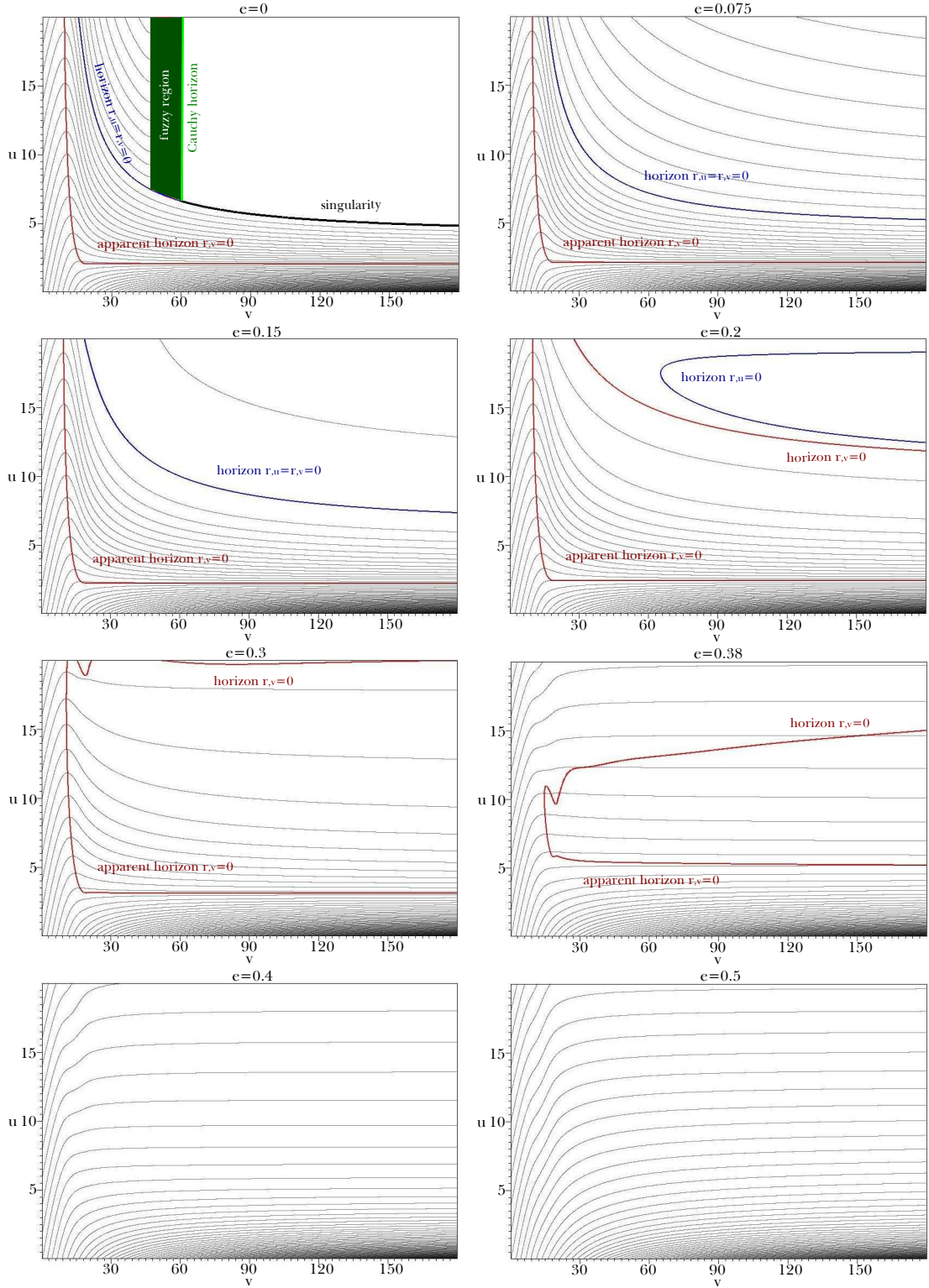}
\caption{\label{fig:P}Numerical results for $P=0.1$ (with Hawking radiation) while varying $e$.}
\end{center}
\end{figure}

We put $P=0.1$ and vary parameters $e$ (Figure~\ref{fig:P}), as was done in the previous section.

For the $e=0$, $0.075$, and $e=0.15$ cases, there is a new horizon where $r_{,u}=r_{,v}=0$. For the $e=0$ case, there is a space-like singularity. Hence, between the $r_{,u}=r_{,v}=0$ horizon and the central singularity, there is a fuzzy region and a Cauchy horizon along an ingoing null direction. The $r_{,v}=0$ and $r_{,u}=0$ horizons can be separated as in the $e=0.2$ case, while for small charge cases, they are mostly foliated. Because $r_{,v}$ and $r_{,u}$ both change their signs around the foliated region, any observer who moves through the horizon sees an increase of areal radius $r$. Therefore, it is a type of a bottleneck in a one-way wormhole. Note that the size of the bottleneck is on the order of $\lesssim P = 0.1$.

For charged black holes in four dimensions, there was a similar effect. However, in those cases, the charge played a crucial role \cite{Hong:2008mw}, while in three dimensions, the charge is not essential to see such a bottleneck, as we can see in the $e=0$ case. Rather, it is the effect due to the negative energy via semi-classical effects that concentrated around the center. This bottleneck becomes a curvature singularity as the curvature functions increase exponentially around these regions. Here, the charge will make the bottleneck relatively regular. This phenomenon will be discussed in Section~\ref{sec:for}.

As we increase the charge, two horizons are separated (in the $e=0.2$ case) and the $r_{,u}=0$ horizon even disappears (in the $e=0.3$ case). This suggests that the amount of the negative energy decreases as the charge increases, approaching and exceeding the extreme limit ($e=0.4$ and $0.5$). In these cases, the inner apparent horizon can be space-like while the outer horizon remains space-like. This implies that the null energy condition is violated only for inside and that it is not violated around the outer horizon. This will be discussed in Section~\ref{sec:ene}.

\subsubsection{\label{sec:for}Formation of weak curvature singularities}

\begin{figure}
\begin{center}
\includegraphics[scale=0.75]{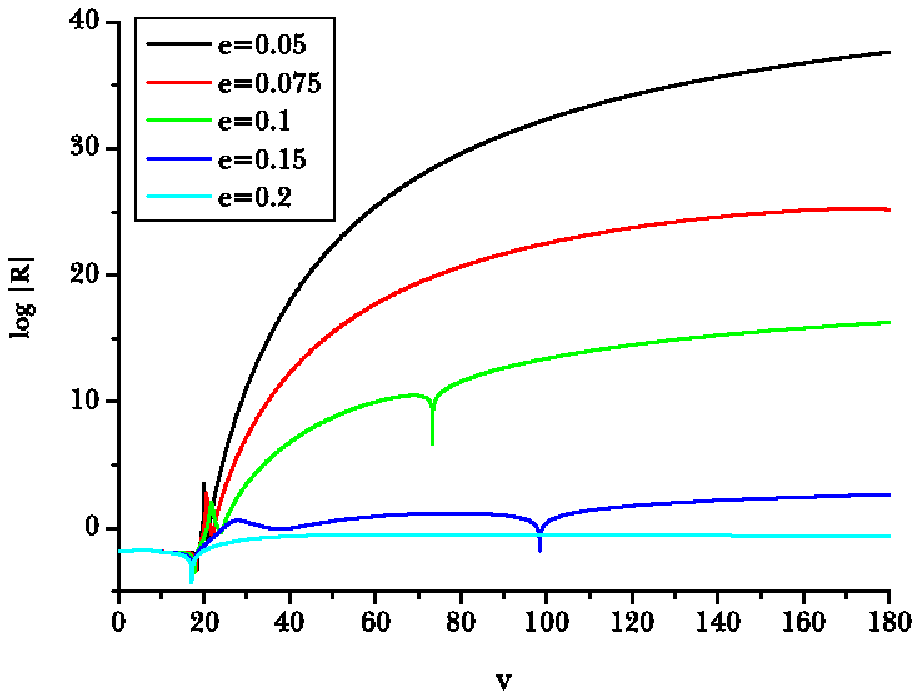}
\includegraphics[scale=0.75]{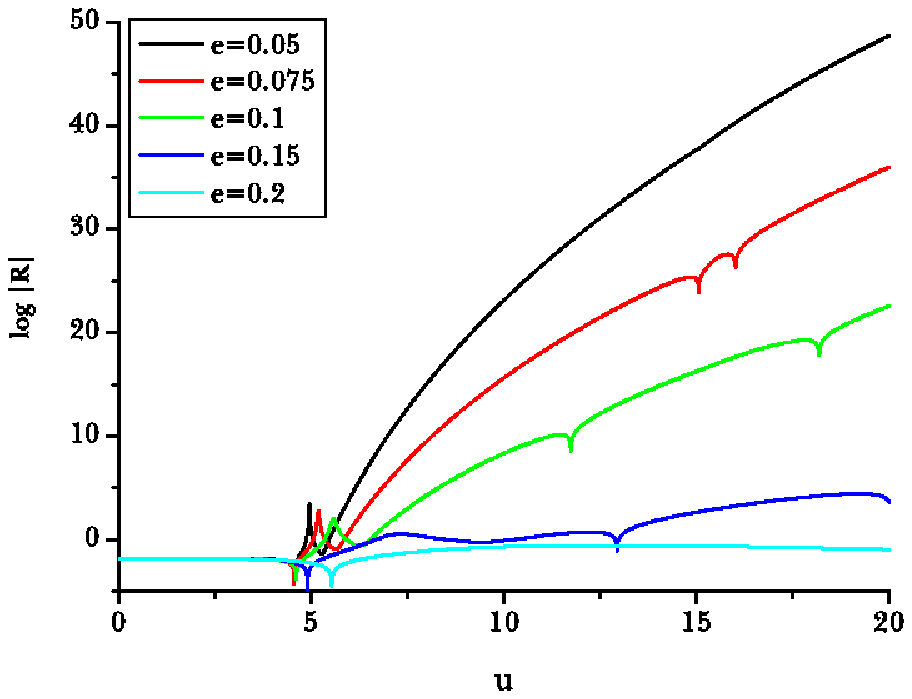}
\caption{\label{fig:Ricci_2}Ricci scalar for $P=0.1$ along $u=15$ when $e=0.05, 0.075, 0.1, 0.5,$ and $0.2$ (Upper) and along $v=180$ when $e=0.05, 0.075, 0.1, 0.5,$ and $0.2$ (Lower).}
\end{center}
\end{figure}

Figure~\ref{fig:Ricci_2} show Ricci scalar functions for $P=0.1$ along $u=15$ and $v=180$ while varying $e$. Until $e=0.15$, the $r_{,v}=0$ horizon and the $r_{,u}=0$ horizon are foliated; hence, the inner horizon becomes a bottleneck of a wormhole, after which the Ricci scalar increases exponentially. Unless we assume a large number of scalar fields $\mathcal{N}$ so that $R < \mathcal{N}$, the internal structure becomes a type of curvature singularity with a non-zero area.

Note that this is not due to mass inflation, because mass inflation does not make the Ricci scalar diverge, as shown in Section~\ref{sec:mas}. Also, this effect exists even for the $e=0$ case. Therefore, we can clarify that this is a new type of curvature singularity.

As was noted in Section~\ref{sec:dim}, the bottleneck occurs via the violation of the null energy condition and the concentration of the negative energy. If the effect of the re-normalized energy-momentum tensor is dominant (around the order of $r \lesssim P$) compared to the classical fields, then the region is no longer semi-classical. Dvali noticed that such a region is not physically well defined and is hence physically problematic. It is interesting that curvature functions become trans-Planckian around such dangerous regions. Such a curvature singularity is found by the authors for the first time, and we will term it a \textit{Dvali curvature singularity} to distinguish it from the mass inflation curvature singularity.

Both types of curvature singularities are weak, in the sense that the area is non-zero while the curvature becomes trans-Planckian \cite{Tipler}. One can compare the order of magnitude of such curvatures between the Dvali curvature singularity and the mass inflation curvature singularity. The former is dominant compared to the latter in terms of the order of magnitude. Therefore, the following conclusions are reasonable:
\begin{itemize}
\item For charged black holes with sufficiently small charge, a space-like singularity partly\footnote{If the bottleneck approaches too closely to the outer horizon, and a space-like singularity can appear, as we see in the $e=0$ case. This is due to the fact that the $T_{vv}$ component is positive around the outer horizon via thermodynamic stability, as we shown in the next section.} changes to a Dvali curvature singularity.
\item For large charged black holes, we can see a mass inflation curvature singularity in the large $v$ limit.
\item Between the two limits, both types of curvature singularities can exist.
\end{itemize}

\subsubsection{\label{sec:ene}Energy-momentum tensors and thermodynamic stability}

\begin{figure}
\begin{center}
\includegraphics[scale=0.25]{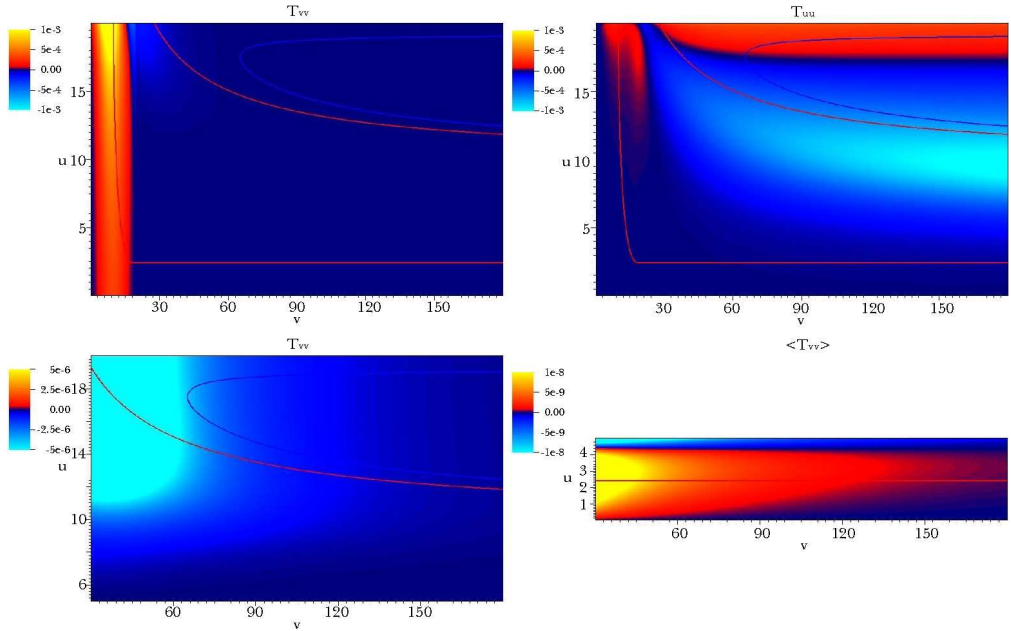}
\caption{\label{fig:energycondition}Energy-momentum tensor components $T_{vv}$, $T_{uu}$, and $T_{vv}$ around the inner horizon, and $\langle \hat{T}^{\mathrm{H}}_{vv} \rangle$ around the outer horizon, for $e=0.2$ and $P=0.1$.}
\end{center}
\end{figure}

Figure~\ref{fig:energycondition} shows the typical distribution of the energy-momentum tensor components of our simulations. The first plot in Figure~\ref{fig:energycondition} is $T_{vv}$, showing that it is positive during the gravitational collapse ($v<20$).

After the gravitational collapse, the inside structure is governed by the negative energy-momentum tensors $T_{uu}$ and $T_{vv}$. The second plot in Figure~\ref{fig:energycondition} shows that $T_{uu}$ is negative around the outer part of the $r_{,u}=0$ horizon and changes its sign around the inner part of the $r_{,u}=0$ horizon. This is consistent given that
\begin{eqnarray}
r_{,uu} |_{r_{,u}=0} = - 8 \pi r T_{uu},
\end{eqnarray}
and hence, if $r_{,u}$ increases along the ingoing null direction at $r_{,u}=0$, $T_{uu}$ should be negative; if $r_{,u}$ decreases along the ingoing direction at $r_{,u}=0$, $T_{uu}$ should be positive.

The third plot in Figure~\ref{fig:energycondition} is $T_{vv}$ around the inner horizon. It is negative around the inner apparent horizon. Note that around the inner horizon,
\begin{eqnarray}
r_{,vv} |_{r_{,v}=0} = - 8 \pi r T_{vv}.
\end{eqnarray}
If $r_{,v}=0$ horizon is inside of the black hole, to increase $r_{,v}$ from negative to positive, the inner horizon should be space-like, and it is equivalent to $T_{vv}<0$. Therefore, this is a consistent behavior.

Finally, the fourth plot in Figure~\ref{fig:energycondition} shows the $\langle \hat{T}^{\mathrm{H}}_{vv} \rangle$ components. Of course, the outer apparent horizon is space-like or time-like if and only if $T_{vv}$ is positive or negative. However, we know that $T_{vv} = T^{\mathrm{C}}_{vv} + \langle \hat{T}^{\mathrm{H}}_{vv} \rangle$ and that competition exists between the two parts. The first classical part is always positive, and hence, to see a time-like horizon, the second quantum part should be negative. Moreover, it has to be greater than the first part. However, the fourth plot shows that the quantum part $\langle \hat{T}^{\mathrm{H}}_{vv} \rangle$ is already positive around the outer horizon. Therefore, there is no hope of seeing the evaporation of a black hole in three-dimensional cases. This is consistent with the known thermodynamic stability of three-dimensional black holes, and this calculation confirms the stability in an independent manner \cite{Reznik:1991qj}.

\subsubsection{Modifications of causal structures due to semi-classical effects}

In this subsection, let us summarize the results of previous sections.

In Section~\ref{sec:cau}, we observed causal structures when there is no Hawking radiation. In Section~\ref{sec:res} and Section~\ref{sec:ene}, we included Hawking radiation and observed some interesting modifications:
\begin{itemize}
\item Space-like singularities are partly changed to a bottleneck of a wormhole (the Dvali curvature singularity).
\item Inner horizons at finite $v$ can be space-like. (This depends on the details of the energy-momentum tensors of inside the black hole.)
\item A $v \rightarrow \infty$ inner horizon remains for sufficiently large charge cases.
\item The outer apparent horizon is thermodynamically stable.
\end{itemize}
According to Section~\ref{sec:mas} and Section~\ref{sec:for}, we understand that the $v \rightarrow \infty$ inner horizons and the bottleneck of the wormhole become curvature singularities. The former is due to mass inflation and the latter is caused by the concentration of negative energy.

\begin{figure}
\begin{center}
\includegraphics[scale=0.6]{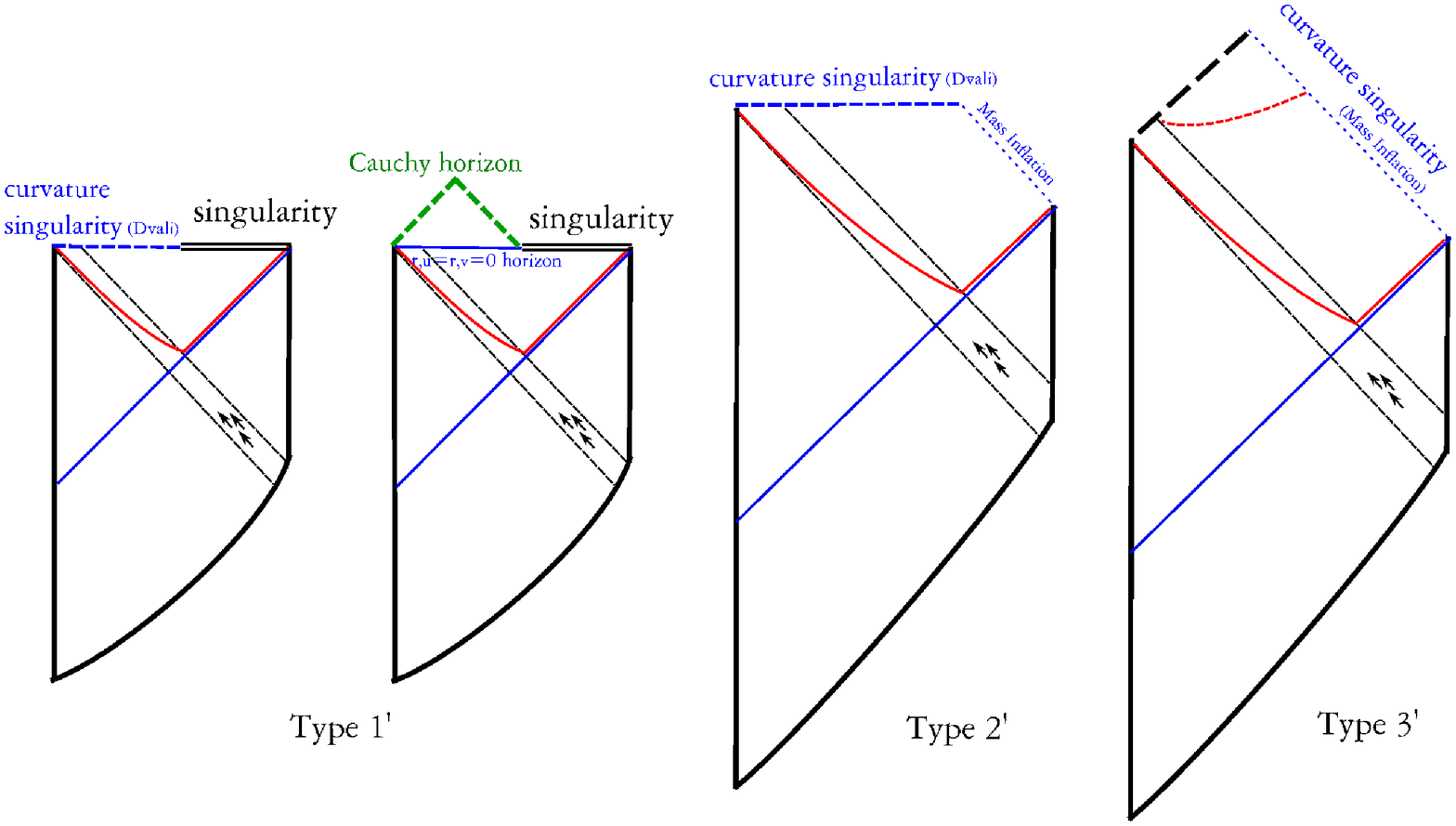}
\caption{\label{fig:P_types}Causal structures with Hawking radiation. Type~$1$, $2$, and $3$ in Figure~\ref{fig:P0_types} should be modified to Type~$1'$, $2'$, and $3'$.}
\end{center}
\end{figure}

Therefore, for Type~$1$ (Figure~\ref{fig:P0_types}), a part of the space-like singularity will be changed by a Dvali curvature singularity (left of Type~$1'$, Figure~\ref{fig:P_types}). We cannot trust beyond the Dvali curvature singularity, since the curvatures become trans-Planckian. However, as was discussed in Section~\ref{sec:dim}, if there are sufficiently large numbers of massless fields that contribute to Hawking radiation without changing $P \propto \mathcal{N} \hbar$, beyond the $r_{,u}=r_{,v}=0$ horizon can be trustable in the semi-classical sense, as the curvatures, for example the Ricci scalar, will decrease in proportion to $1/\mathcal{N}^{2}$. If this is true, our solution implies a branching-off geometry (right of Type~$1'$, Figure~\ref{fig:P_types}). For Type~$2$ (Figure~\ref{fig:P0_types}), again, the space-like singularity should be changed by a Dvali curvature singularity and the inner Cauchy horizon should be changed by the mass inflation singularity (Type~$2'$, Figure~\ref{fig:P_types}). For Type~$3$ (Figure~\ref{fig:P0_types}), the semi-classical effect can modify the inner horizon to be space-like and there will be a mass inflation singularity in the $v \rightarrow \infty$ limit (Type~$3'$, Figure~\ref{fig:P_types}). Of course, in that limit, the out-going Cauchy horizon will be also a type of mass inflation singularity. In this limit, we may not see the Dvali curvature singularity.

\section{\label{sec:con}Conclusion}

In this paper, we investigated the dynamical formation and evolution of three-dimensional charged black holes. As a concluding remark, we summarize our new results and point out some interesting issues in terms of two points: classification of causal structures for dynamical black holes and cosmic censorship in three dimensions.

\subsection{Causal structures}

We study gravitational collapses of charged matter fields and classified causal structures (Figure~\ref{fig:P0_types}): Type~$1$ for an almost neutral case, Type~$2$ for a small charge, Type~$3$ for a large charge, and Type~$4$-$1$ or Type~$4$-$2$ for an excessive charge. For near-extreme cases, we show evidence that Type~$4$-$2$ will be more probable. When there is an inner Cauchy horizon in $v \rightarrow \infty$ limit, a local energy density for an in-going null observer increases exponentially, and this confirms mass inflation. This will eventually create a curvature singularity in a large $v$ or large $u$ limit. However, this was significantly different from the four-dimensional case, as in those cases, other types of curvature functions also diverge.

As we include semi-classical effects via Hawking radiation, we note that the outer apparent horizon is not affected and is thermodynamically stable. However, the internal structure will be significantly modified. Negative energy is concentrated near the center, and this can create a bottleneck in a wormhole. Such a bottleneck is not due to a charge but is caused by negative energy. This neck occurs when the semi-classical effects dominate the classical effects. We observed that the region around the bottleneck becomes a curvature singularity (the Dvali curvature singularity). Moreover, the inner horizon can be space-like via negative energy. Therefore, we delineate two types of curvature singularities in three dimensions: the Dvali curvature singularity and the mass inflation curvature singularity. Therefore, as we summarize these contents, the causal structures Type~$1$, Type~$2$, and Type~$3$ should be modified to Type~$1'$, Type~$2'$, and Type~$3'$ (Figure~\ref{fig:P_types}).

\subsection{Comments on cosmic censorship}

One interesting finding was related to cosmic censorship. There is some controversy about whether or not there is a violation of cosmic censorship in anti de Sitter space \cite{Hertog:2003zs}\cite{Gutperle:2004jn}. In terms of BTZ solutions, there appears to be a time-like or null singularity for $M < Q^{2} \ln Q^{2}$ or $M=Q=0$ cases. Therefore, as long as we rely on the BTZ solution, a violation of weak cosmic censorship is not odd.

However, if we believe that all realistic black holes should be formed by a gravitational collapse, our conclusions seem to say that there is no violation of weak cosmic censorship (Section~\ref{sec:cri}). When there is no horizon and excessive charge, the tendency of the charge distribution shows that the energy density around the center is regular.

The absence of a naked singularity has also been observed in four-dimensional gravitational collapses \cite{doublenull}\cite{Hong:2008mw}. However, in that case, as we give an excessive charge, a near-extreme black hole was formed rather than a disappearance of a black hole. This is an interesting difference between three dimensions and four dimensions.

In terms of strong cosmic censorship, there can be two ways to touch the Cauchy horizon. First, there is the Cauchy horizon along the in-going null direction (e.g., the thin red dotted line in Type~$3$ in Figure~\ref{fig:P0_types}). Second, there is the Cauchy horizon along the out-going null direction (e.g., thick black dashed line in Type~$3$ in Figure~\ref{fig:P0_types}). Given that this black hole does not evaporate, for the former Cauchy horizon, an ingoing observer requires infinite advanced time $v$ (although this does not imply infinite proper time). Thus, via mass inflation, it will surely be a curvature singularity. Therefore, it is likely not possible to touch in the semi-classical regime. Of course, it may be possible for an observer to touch the latter Cauchy horizon. However, we know that when there are two horizons (a space-like outer horizon and a time-like inner horizon), it is not evident whether there is a time-like singularity for an ingoing null observer, as we noted in Table~\ref{table:fitting}. If this is not the case, then strong cosmic censorship appears to be safe.

\section*{Acknowledgment}
The authors would like to thank Bum-Hoon Lee, Ewan Stewart, and Seungjoon Hyun for discussions and encouragement. DY and DH were supported by Korea Research Foundation grants (KRF-313-2007-C00164, KRF-341-2007-C00010) funded by the Korean government (MOEHRD) and BK21. DY was supported by the National Research Foundation of Korea(NRF) grant funded by the Korea government(MEST) through the Center for Quantum Spacetime(CQUeST) of Sogang University with grant number 2005-0049409. HK was supported by the National Research Foundation of Korea(NRF) grant funded by the Korea government(MEST) with the grant number 2009-0074518.

\section*{\label{sec:appa}Appendix A. The Birkhoff theorem for charged BTZ black holes}

In this appendix, we briefly examine the Birkhoff theorem for charged BTZ black holes. There is an argument that the only spherical symmetric solution of $(2+1)$-dimensional space-time with a negative cosmological constant is a BTZ black hole or the self-dual Coussaert-Henneaux space-time \cite{Birkhoff}. This was applied only for neutral cases, and hence it is necessary to present an argument about the uniqueness of the solution when the electric charges are involved. In this appendix, we prove that any spherical symmetric charged black holes in a three-dimensional anti de Sitter background will approach the charged BTZ asymptotically.

The metric of a spherically symmetric space-time can always be cast in the form
\begin{eqnarray}
ds^{2} = - e^{2\psi(r,t)}f(r,t) dt^2 + \frac{1}{f(r,t)} dr^2 + r^2 d\varphi^2
\end{eqnarray}
involving the two arbitrary functions $\psi(r,t)$ and $f(r,t)$.
We put the gauge field by
\begin{eqnarray}
A_{t} (r) = Q \ln \frac{r}{l}, \label{gauge}
\end{eqnarray}
as the electric field strength must be of the order $1/r$ asymptotically. We may put the gauge field as an arbitrary function for completeness. However, this restriction of the configuration space is reasonable because the conservation of the energy-momentum tensor, $\nabla_\mu {T^\mu}_\nu = 0$, constrains the form of the gauge field as Equation~(\ref{gauge}).

The independent Einstein equations are as follows:
\begin{eqnarray}
-\frac{1}{l^2} + \frac{Q^2}{r^2}e^{-2\psi} + \frac{{f'}^2}{2r} = 0, \quad && (tt)\mathrm{-component} \\
\frac{1}{2r f^2}e^{-2\psi}\dot{f} = 0, \quad && (tr)\mathrm{-component} \\
-\frac{1}{l^2} + \frac{Q^2}{r^2}e^{-2\psi} + \frac{{f'}^2}{2r} + \frac{f}{r}\psi' = 0, \quad && (rr)\mathrm{-component}
\end{eqnarray}
where $'$ and $\dot{}$ represent the derivatives with respect to $r$ and $t$, respectively.

From the $(rr)$-component and the $(tt)$-component of the Einstein equations, we have $\psi = \psi(t)$. In addition, a time redefinition can absorb $\psi(t)$ and it will make $f(r,t)$ be the only function to be determined.
From the $(tr)$-component, we have $f=f(r)$.
Finally, from the $(tt)$-component of the Einstein equations, we obtain
\begin{eqnarray}
f(r) = \frac{r^2}{l^2} + C - Q^2 \ln \left( \frac{r^2}{l^2} \right),
\end{eqnarray}
where $C$ is the integration constant and we can choose by $-M$. Therefore, this proves our assertion. Any spherical symmetric charged black holes in the three-dimensional anti de Sitter background will approach the charged BTZ solution.

\begin{figure}
\begin{center}
\includegraphics[scale=0.75]{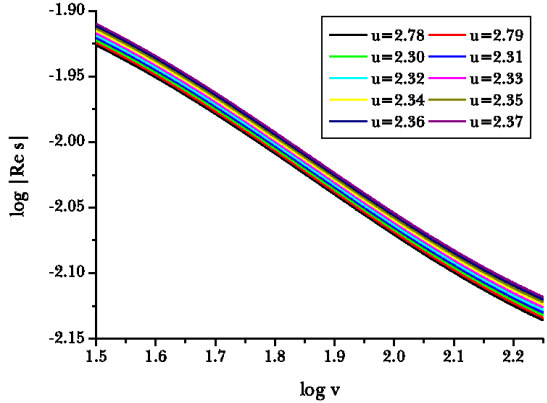}
\caption{\label{fig:hair}The real part of the scalar field as a function of $v$ around the outer apparent horizon. Here, we used $e=0.2$ and $P=0$. The gradients of each lines are approximately $-0.29$.}
\end{center}
\end{figure}

To sure this expectation, we test the dynamics of scalar field around the outer apparent horizon. For $e=0.2$ and $P=0$, we observe the behavior of the scalar field (Figure~\ref{fig:hair}). For convenience, we observed the real part of the scalar field as a function of $v$ around the outer apparent horizon ($2.78\leq u \leq2.87$); the imaginary part is also qualitatively similar. The scalar field falls off approximately $\sim v^{-p}$ with $p \simeq 0.29$. Therefore, in the $v \rightarrow \infty$ limit, the scalar field will decay to zero. Therefore, this makes sure that our solutions will emerge to a vacuum solution in the anti de Sitter. Then, our previous proof shows that our results surely emerge to BTZ solutions.

\section*{\label{sec:appb}Appendix B. Consistency and convergence tests}

Now we test consistency and convergence of this paper.

\begin{figure}
\begin{center}
\includegraphics[scale=0.75]{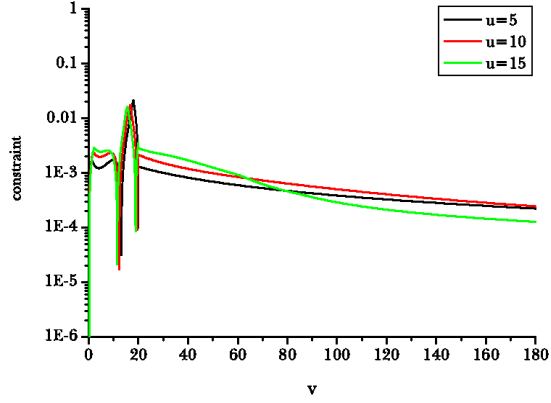}
\includegraphics[scale=0.75]{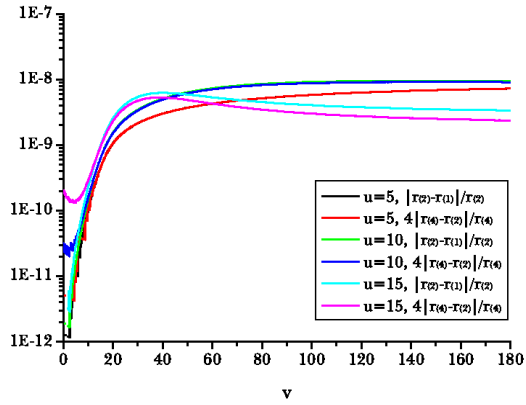}
\includegraphics[scale=0.75]{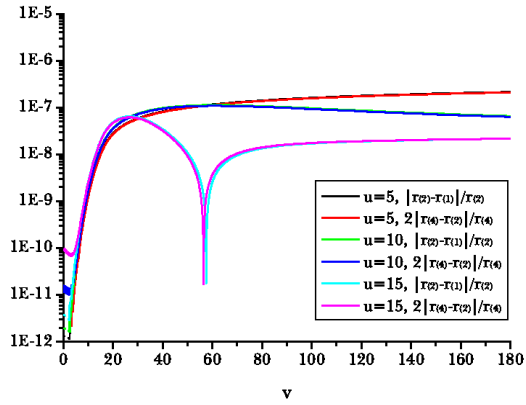}
\caption{\label{fig:convergence}Top: constraint function $|g_{,v} - 2 g d + 2rz \bar{z} + 4P(d_{,v}-d^{2})|/(|g_{,v}| + |2 g d| + |2rz \bar{z}| + 4P(|d_{,v}|+|d^{2}|))$ for $P=0.1$ and $e=0.2$ along $u=5$, $10$, and $15$. Middle and Bottom: convergence tests $|r_{(2)}-r_{(1)}|/r_{(2)}$ and $2|r_{(4)}-r_{(2)}|/r_{(4)}$ for $P=0$ and $P=0.1$ with $e=0.2$ along $u=5$, $10$, and $15$, where $r_{n}$ means an $n \times n$ finer simulation.}
\end{center}
\end{figure}

First, to check the consistency, we used one of constraint equations. As an example, we choose the constraint function Equation~(\ref{eq:E2})
\begin{eqnarray}
\frac{|g_{,v} - 2 g d + 2rz \bar{z} + 4P(d_{,v}-d^{2})|}{|g_{,v}| + |2 g d| + |2rz \bar{z}| + 4P(|d_{,v}|+|d^{2}|)}
\end{eqnarray}
to see the constraint. The value should be exactly zero, but due to numerical errors, it may not be zero. The normal scale of the constraint function is on the order of $1$; hence, if the value is sufficiently less than $1$, it is certain that our simulations give a sound result. Top in Figure~\ref{fig:convergence} shows the constraint function for the $P=0.1$ and $e=0.2$ case along the $u=5$, $10$, and $15$ slices. The result is sufficiently small and the order is less than $1 \%$. Note that when the constraint becomes on the order of $1 \%$, the $r_{,vv}$ actually approaches zero and the behavior of the constraint function becomes unstable. However, this is not a significant problem for entire causal structures.

Second, to check the convergence, we first see $|r_{(2)}-r_{(1)}|/r_{(2)}$ and $4|r_{(4)}-r_{(2)}|/r_{(4)}$ along the $u=5$, $10$, and $15$ slices, where $r_{n}$ means an $n \times n$ finer simulation (Middle in Figure~\ref{fig:convergence}). We checked for the $e=0.2$ and $P=0$ case. For each slice, two curves are almost same; this shows the second order convergence. Numerical errors are less than $0.000001 \%$. Next, we test the $e=0.2$ and $P=0.1$ case (Bottom in Figure~\ref{fig:convergence}) and compare $|r_{(2)}-r_{(1)}|/r_{(2)}$ and $2|r_{(4)}-r_{(2)}|/r_{(4)}$. As we noted, it shows the first order convergence. However, the error is sufficiently small again, and hence it is reasonable to trust our numerical calculations.

\end{document}